\documentclass[prc,preprint,
    floatfix,
    tightenlines,
    amsfonts,nofootinbib,superscriptaddress]{revtex4-1}

\pdfoutput=1

\usepackage{graphicx}  %
\usepackage{bm}  %
\usepackage{amssymb}
\usepackage{dcolumn}

\usepackage[normalem]{ulem}
\usepackage{amsmath}
\usepackage{subfigure}

\graphicspath{{./PicsData/}{./PicsData1D/}}

\begin{document}

%

% general definitions
\newcommand{\beqn}{\begin{equation}}
\newcommand{\eeqn}{\end{equation}}
\newcommand{\bea}{\begin{eqnarray}}
\newcommand{\eea}{\end{eqnarray}}

\newcommand{\vlowk}{V_{{\rm low}\,k}}
\newcommand{\vsrg}{V_s}
\newcommand{\tlowk}{T_{{\rm low}\,k}}
\newcommand{\trel}{T_{\rm rel}}
\newcommand{\vnn}{V_{\rm NN}}

\newcommand{\fm}{\, \text{fm}}
\newcommand{\fmi}{\, \text{fm}^{-1}}
\newcommand{\mev}{\, \text{MeV}}

\newcommand{\la}{\langle}
\newcommand{\ra}{\rangle}
\newcommand{\ts}{\textstyle}
\newcommand{\wt}{\tilde}
\newcommand{\wh}{\widehat}

\newcommand{\qvec}{{\bf q}}
\newcommand{\kvec}{{\bf k}}
\newcommand{\kpvec}{{\bf k'}}
\newcommand{\kmax}{k_{\rm max}}

\newcommand{\adag}{a^\dagger}
\newcommand{\adaggera}{a^\dagger_{q} a^{{\protect\phantom\dagger}}_{q}}
\newcommand{\adaggeraop}{(\adaggera)_s}
\newcommand{\Hzero}{H^{\rm bd}}
\newcommand{\flow}{s}
\newcommand{\oneoverr}{r^{-1}}

\newcommand{\mystrut}{\rule[-1mm]{0mm}{5.5mm}}

\newcommand{\Oop}{\wh{O}}
\newcommand{\projP}{\mathcal{P}}
\newcommand{\projQ}{\mathcal{Q}}

%
%
%
% End: Simple substitution macros
%

% uncomment \draft to have PACS numbers appear

%
\title{Operator  Evolution via the \\Similarity Renormalization Group I:\\ The Deuteron}

\author{E.R.\ Anderson} \email{anderson@physics.ohio-state.edu}
\affiliation{Department of Physics, The Ohio State University, Columbus, OH 43210}
\author{S.K.\ Bogner}\email{bogner@nscl.msu.edu}
\affiliation{National Superconducting Cyclotron Laboratory and Department of Physics and Astronomy, Michigan State University, East Lansing, MI 48824}
\author{R.J.\ Furnstahl}\email{furnstahl.1@osu.edu}
\affiliation{Department of Physics, The Ohio State University, Columbus, OH 43210}
%\author{E.D.\ Jurgenson}\email{jurgenson.1@osu.edu}
%\affiliation{Department of Physics, The Ohio State University, Columbus, OH 43210}
\author{R.J.\ Perry}\email{perry@mps.ohio-state.edu}
\affiliation{Department of Physics, The Ohio State University, Columbus, OH 43210}

\date{\today}

\begin{abstract}
%%%%%%%%%%%%%%%%%%%%%%%%%%%%%%%%
%
Similarity renormalization
group (SRG) flow equations can be used to unitarily soften nuclear
Hamiltonians by decoupling high-energy intermediate state contributions to low-energy observables
while maintaining the natural hierarchy of many-body forces.
Analogous flow equations can be used to
consistently evolve operators so that observables
are unchanged if no approximations are made.  The question in practice
is whether the advantages of a softer Hamiltonian and less correlated wave
functions might be offset by complications in approximating and
applying other operators.
Here we examine the properties of SRG-evolved operators,  
focusing in this article on applications to the 
deuteron but leading toward methods for few-body systems.
We find the advantageous features generally carry over to other operators 
with additional simplifications in some cases
from factorization of the unitary transformation operator.
\end{abstract}

\maketitle

%%%%%%%%%%%%%%%%%%%%%%%% Introduction %%%%%%%%%%%%%%%%%%%%%%%%%%%%%

\section{Introduction} \label{sec:intro}

Renormalization group methods can be used to soften interactions
in nuclear systems, which extends the range of many
computational methods and qualitatively improves their convergence
patterns~\cite{Bogner:2009bt}.
The similarity renormalization group (SRG) \cite{Glazek:1993rc,Wegner:1994,Szpigel:2000xj,Glazek:2001uw,Bogner:2006pc}
does this by systematically evolving 
Hamiltonians via a continuous series of unitary transformations 
chosen to decouple the
high- and low-energy matrix elements of a given interaction
\cite{Jurgenson:2007td,Jurgenson:2009qs}.
In particular, a flow equation with parameter $s$
and generator $\eta_s\equiv [G_{s},H_{s}]$, 
\beqn
  \frac{dH_{s}}{ds}=[\eta_s,H_{s}] 
   \;,
   \label{eq:Hflow}
\eeqn
unitarily evolves an initial Hamiltonian $H_{s=0} \equiv H=\trel+V$. 
Choosing the flow operator $G_{s}$ specifies the SRG evolution. 
This equation implements the unitary transformation
\beqn
  H_{s}=U_{s}H_{s=0}U^{\dagger}_{s}=\trel+V_{s}
  \;,
  \label{eqn:first}
\eeqn
which defines $V_s$ by choosing the relative
kinetic energy to be invariant, and 
where the generator $\eta_s$ is related to $U_s$ by
\beqn
  \eta_{s}=\frac{dU_{s}}{ds}U^{\dagger}_{s}=-\eta^{\dagger}_{s}
  \;.
  \label{eqn:generator}
\eeqn
As in most previous nuclear applications, 
here we take $G_{s}=\trel$ to
suppress off-diagonal elements of the Hamiltonian in momentum space
(see Sec.~\ref{sec:DeuteronExpectVal}; alternative choices are
discussed in Refs.~\cite{Anderson:2008mu} and \cite{Anderson:2010aa}).
This decoupling leads to greatly improved convergence of the binding energy 
in few and
many-body calculations~\cite{Bogner:2007rx,Jurgenson:2009qs}.  
A major advantage of the SRG relative to other energy-independent renormalization group (RG) methods is that the Hamiltonian flow equation is formulated solely in terms of the evolving Hamiltonian and does not involve the T-matrix, which avoids issues with solving equations in multiple
channels and allows any convenient basis to be used;
these features mean that evolving few-body forces 
is practical~\cite{Jurgenson:2008jp,Jurgenson:2009qs}.

To use the wave functions produced by 
SRG-evolved interactions to calculate other matrix elements of
interest, we cannot in general neglect the associated change in operators.   
The evolution of any operator $\Oop\equiv\Oop_{s=0}$ is given by the same
unitary transformation used to evolve the Hamiltonian~\cite{Szpigel:2000xj,Bogner:2007jb},
\beqn
  \Oop_{s}=U_{s}\Oop_{s=0}U^{\dagger}_{s}
  \;,
   \label{eq:Otransform}
\eeqn
which implies by differentiation with respect to $s$
the general operator SRG equation
\beqn
\frac{d\Oop_{s}}{ds}=[\eta_{s},\Oop_{s}]
  \;.
  \label{eq:Opevolve}
\eeqn
Although this equation can be used to find $\Oop_s$, it
is computationally efficient to construct the unitary
transformation directly from the eigenvectors of the evolved and
unevolved Hamiltonian using   
\beqn
U_{s}=\sum_{\alpha}\left\vert \psi_{\alpha}(s)\right\rangle\left\langle  \psi_{\alpha}(0) \right\vert
  \;, 
  \label{U:basictrans}
\eeqn
where the sum on $\alpha$ is over all eigenvectors,
and then to apply Eq.~\eqref{eq:Otransform} directly. 
In practice we work in a discretized basis so this sum is
finite and Eq.~\eqref{eq:Otransform} is a simple matrix
product.
In cases where both methods have been used to
calculate the SRG evolution, the transformations produced by
Eqs.~(\ref{eq:Opevolve}) and (\ref{U:basictrans}) agree up to
numerical errors.  However, it has been found that the
one-step transformation produced from the eigenvectors is numerically
more robust than the differential equation. The direct construction of
\(U_{s}\) via Eq.~(\ref{U:basictrans}) is used to
calculate operator evolution throughout this work.

If implemented without approximation, unitary transformations
preserve operator matrix elements by construction,
\beqn
  \langle \psi_\alpha(s) | \Oop_s | \psi_{\alpha'}(s) \rangle
  = 
  \langle \psi_\alpha(0) | \Oop_{s=0} | \psi_{\alpha'}(0) \rangle
  \;,
  \label{eq:equivalent}
\eeqn
and thus preserve the physics in the initial Hamiltonian and other operators.
But do the advantages of the SRG evolution of Hamiltonians carry over
to other operators and are there problems with the practical implementation
of operator evolution?  
Equation~\eqref{eq:equivalent} implies that 
changes in the wave function are ``compensated'' by changes in
the operator.  This might reasonably be expected to shift around
the physics while conserving the computational difficulty, so that at best we
must deal with either a simple operator and complicated wave functions
or a complicated operator and simplified wave functions.
However,
the SRG evolution of the Hamiltonian generates a much simpler interaction
(smoother and decoupled), which leads to a simpler
wave function (reduced short-range correlations).%
\footnote{Note that the interpretation of ``simpler'' can vary.
For some Monte Carlo methods,
the SRG Hamiltonians become more
complicated in the sense of increasing non-locality in coordinate representation.}
What about other operators?
Could there be strong and/or fine-tuned cancellations between the evolved
wave functions and evolved operators?  Can the hierarchy
of many-body contributions be violated?
We address these questions in this article and
a sequel~\cite{Anderson:2010aa}.

The evolution of three-body and higher-body interactions is critically
important for the SRG and a parallel discussion is needed for other
operators.
To see how
one-, two-, three-, and higher-body operators can be identified, it is useful 
to decompose the running SRG operator $\Oop_{s}$ in
second-quantized form.  Schematically (suppressing indices and sums),
\beqn
\Oop_{s} = \la \Oop_{s}^{(1)} \ra \, \adag a  
+ \la \Oop_{s}^{(2)} \ra \, \adag \adag a a
+ \la \Oop_{s}^{(3)} \ra \, \adag \adag \adag a a a + \ldots \,,
\label{eq:2ndquant}
\eeqn
where $\adag$ and $a$ are creation and annihilation operators with
respect to the vacuum in a single-particle basis.  This \emph{defines}
$\la \Oop_{s}^{(1)} \ra$, $\la \Oop_{s}^{(2)} \ra$, 
$\la \Oop_{s}^{(3)} \ra$,
. . .\ as the one-body, two-body, three-body, . . .\
operator matrix elements in that basis at each ${s}$.  
The SRG evolution in Eqs.~\eqref{eq:Hflow} and \eqref{eq:Opevolve}
is dictated by commutators involving $\Oop_{s}$ and $H_{s}$
(which also has such an expansion).
When they are evaluated, we see that even if initially there are only
one-body operators, higher-body terms will appear in both  $\Oop_{s}$ and $H_{s}$
 with each
step in ${s}$.  Thus, when applied in an $A$-body subspace, the SRG
will ``induce'' $A$-body operators.  
However, we find that each $\la \Oop_{s}^{(n)} \ra$ is determined
fully in the $A=n$ subspace, with no dependence on higher-body operators.
This allows us to extract the induced many-body components of the
operator as needed~\cite{Anderson:2010aa}.
Note that for Hamiltonians with no external potentials (i.e., no one-body interactions),
$\la \Oop_{s}^{(1)} \ra$ is
independent of $s$~\cite{Anderson:2010aa}. 
It is also important to remember that input operators, \(\Oop_{s=0}\), for low-energy
effective theories are generally never simply one- or two-body
operators, although these components may dominate.
The question is whether an initial many-body hierarchy (expected from chiral effective field theory formulations of the nuclear interaction) is maintained
by the SRG evolution.

To avoid confusion, we note that it is
also possible to normal order with respect to a finite-density reference state instead of the 
vacuum, which leads to the ``in-medium SRG'' (see Ref.~\cite{Tsukiyama:2010rj} 
and references therein).
This changes the
definition of the matrix elements and
creation and annihilation operators in Eq.~\eqref{eq:2ndquant}
and shifts higher-body pieces to the zero-body, one-body, and two-body levels.
These operators
are well defined but have different properties from those
considered here; for example, even one-body operators flow.
The in-medium SRG shows great promise as a microscopic method
of deriving effective shell model interactions for nuclei and
the study of the corresponding operators is an important topic for 
future investigation. 

In this article, we restrict our attention to the deuteron,
which means only one- or two-body operators are relevant.
Furthermore,
any running with $s$ is due to an induced two-body part.
In practice, working in the 
two-body system in the center-of-mass frame makes the occurrence
of two-body operators uneventful; there are far
more consequences for $A>2$.
We defer to the sequel~\cite{Anderson:2010aa} 
the discussion of evolving and extracting operator components
and embedding them in larger $A$ systems using a harmonic
oscillator basis.

Nuclear SRG studies to date have
focused primarily on the calculation of binding energies and phase shifts to analyze the characteristics of the SRG-evolved
interactions.  A limited analysis of the deuteron momentum
distribution defined by the initial operator $\adaggera$
has also been made \cite{Bogner:2007jb} and 
decoupling of long-distance operators for the rms radius,
quadrupole moment, and $\oneoverr$ have been examined~\cite{Jurgenson:2007td},
but only for the bare operators.
We expand upon these studies here, 
emphasizing the nature of SRG-evolved operators
and also include the first calculations of deuteron charge,
quadrupole, and magnetic form factors.
However, we use these electromagnetic operators as test cases for
addressing  questions about operator evolution and not yet for 
systematic comparison to experiment, which requires a more complete treatment of the \(s=0\) operators.

Other issues arise about processes with large momentum
transfers, such as $(e,e'p)$.  Theoretical analyses relate such
experiments to nuclear momentum distributions if
the impulse approximation is assumed to be valid for a high-cutoff interaction \cite{Frankfurt:2008zv}.
Calculations find
nearly universal scaling of the high-momentum tails, which is 
interpreted in terms of short-range correlations in the nuclear wave functions.
It might be thought naively that this physics is beyond the reach
of low-momentum approaches, for which wave functions have drastically
reduced short-range correlations.  However, Eq.~\eqref{eq:equivalent}
is unequivocal:  The experimental cross section is unchanged with
SRG evolution to low momentum if no approximations are made, even if the evolved wave function has almost no short range correlations.  
But does the calculation become intractable because the evolution
of the momentum occupation operator makes it too
complicated (e.g., strong non-localities, too-large many-body components)?
We begin to address this question by showing that under the relevant
kinematic conditions there is \emph{factorization} of the unitary
transformation $U_s$, which leads to significant simplifications and
an alternative interpretation of the universal high-momentum dependence and scaling.

The plan of the article is as follows. 
In Sec.~\ref{sec:DeuteronExpectVal}, we explore whether decoupling
of the Hamiltonian for two-body systems is mirrored in the operator flow, 
focusing on evolution of the momentum distribution  as a
characteristic example.  
In Sec.~\ref{sec:otherops}, we consider the evolution
of other operators, including electromagnetic form factors
in the deuteron, which
are all found to flow to smooth, low-momentum forms.
Variational calculations of the evolved deuteron binding energy and operator matrix elements are explored in Sec.~\ref{sec:variational}. The factorization of the unitary transformation operator under certain conditions is demonstrated
in Sec.~\ref{sec:opfactor}, along with a
model calculation of the momentum distribution for $A\geq 2$.
Our conclusions are summarized in Sec.~\ref{sec:summary}.

%%%%%%%%%%%%%%%%%%%%%%%%%%%%%%%%%%%%%%%%%%%%%%%%%%%%%%%%%%%%%%%%%%%%%%%5
\section{Decoupling and the Deuteron Momentum Distribution  
          \label{sec:DeuteronExpectVal}}

\subsection{Potentials and Decoupling}

In this section, we specialize the SRG evolution to a two-particle
partial-wave momentum basis with flow operator $G_s = \trel$.  
The flow equation, 
\beqn
  \frac{dH_{s}}{ds}=[[\trel,H_{s}],H_{s}] = \frac{dV_{s}}{ds} \;,
\eeqn
(recall that \(\trel\) is  chosen to remain constant)
for nucleon-nucleon ($NN$) potentials is projected onto each channel using
\(1=\frac{2}{\pi}\int^{\infty}_{0}q^{2}dq\left\vert q
\right\rangle\left\langle q \right\vert \) with
$\hbar = M =1$, yielding 
\bea
  \frac{dV_{s}(k,k')}{ds} &=&
     -(k^{2}-k'^{2})^{2}\,V_{s}(k,k')
     \nonumber \\
     & &  \null + \frac{2}{\pi} \int^{\infty}_{0}\! q^{2}dq\, 
     (k^{2}+k'^{2}-2q^{2}) V_{s}(k,q) V_{s}(q,k')
     \;.
     \label{eq:dVdskkp}
\eea
This equation is implemented in a discretized Gaussian quadrature basis as a
set of coupled differential equations for the matrix elements 
(angular momentum indices in the coupled channels have been suppressed)
that are solved numerically.

\begin{figure}[tbh]
\subfigure[]{\includegraphics[width=0.92\textwidth]{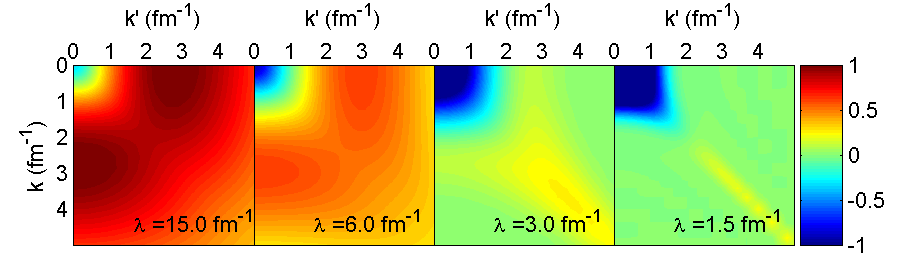}}
\subfigure[]{\includegraphics[width=0.92\textwidth]{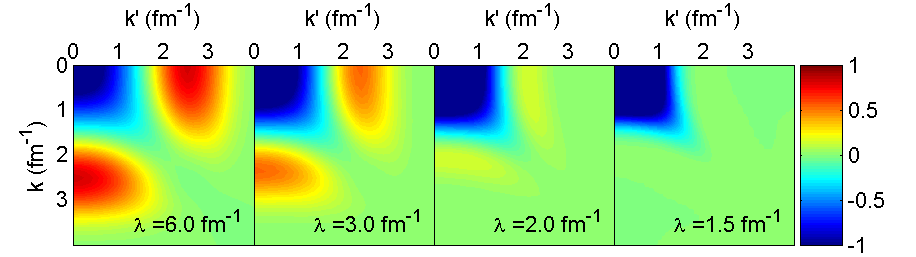}}
 \caption{(Color online) SRG evolution of the momentum-space
 $^3$S$_1$ potential starting
 with (a) Argonne $v_{18}$ (AV18)~from \(\lambda=15 \fmi\) to \(\lambda=1.5\fmi\) \cite{Wiringa:1994wb} and (b)
 N$^3$LO (500\,MeV)~from \(\lambda=6 \fmi\) to \(\lambda=1.5\fmi\) \cite{Entem:2003ft}.}
 \label{fig:potential}
\end{figure}

The flow is visualized in Fig.~\ref{fig:potential} for two representative
initial potentials, Argonne $v_{18}$ (AV18)~\cite{Wiringa:1994wb}
and a chiral N$^3$LO (500\,MeV) effective field theory (EFT)
potential from Ref.~\cite{Entem:2003ft},
which are currently the most commonly used $NN$ interactions for microscopic
nuclear structure calculations.
The figures show snapshots of the potential matrix $V_s(k,k')$
in the $^3$S$_1$ partial wave
at several values of $\lambda$, which is a useful alternative
parameter to characterize the flow. It is related to $s$ by $\lambda = s^{-1/4}$.
The scale at the right indicates the strength of the potential
(in fm); Gaussian mesh weights are not shown.  
One can see that the large matrix elements between low
and high momentum at
the initial $\lambda$'s shown (particularly for AV18)
are suppressed by $\lambda=1.5\fmi$; that is, we see decoupling for both potentials.

The decoupling seen in Fig.~\ref{fig:potential} is readily understood
from the SRG flow equations.
Because of the dominance of the kinetic
energy, Eq.~\eqref{eq:dVdskkp} 
for sufficiently off-diagonal $k$ and $k'$ is given to
good approximation by
\beqn
  \frac{dV_{s}(k,k')}{ds}\approx-(k^{2}-k'^{2})^{2}V_{s}(k,k')
  \;,
\eeqn 
which, when solved, predicts 
\beqn
  V_{s}(k,k')\approx e^{-s(k^{2}-k'^{2})^{2}}\, V_{s=0}(k,k')=e^{-\frac{(k^{2}-k'^{2})^{2}}{\lambda^{4}}}\, V_{\lambda=\infty}(k,k')
   \;\label{eqn:ApproxDecoupling}.
\eeqn
Thus, using this generator we can see that the far off-diagonal elements
of the potential matrix are suppressed exponentially with an
approximate width given by the flow parameter \(\lambda\)
\cite{Bogner:2006pc}.
(If the potential is plotted as a function of $k^2$, the width
of the partially diagonalized potential is clearly seen to be well approximated
by $\lambda^2$~\cite{Bogner:2009bt}.)
While it is not evident from the figure, the potential
is also very smooth.

It is not immediately clear, however, that this decoupling will be
advantageous for the calculation of observables other than the
binding energy. 
If we project the operator flow equation
\beqn
  \frac{d\Oop_{s}}{ds}=[[\trel,H_{s}],\Oop_{s}]
  \;,
\eeqn
onto the partial wave momentum basis, we find 
\beqn
 \frac{dO_{s}(k,k')}{ds} =
    \frac{2}{\pi}\int^{\infty}_{0}\! q^{2}dq\,
    [(k^{2}-q^{2})V_{s}(k,q)O_{s}(q,k')
     + (k'^{2}-q^{2})O_{s}(k,q)V_{s}(q,k')]
  \;,
 \label{O:DEevolve}
\eeqn  
and decoupling is not manifest.
Further, it is not clear if \(\lambda\) provides a measure
of decoupling in the case of a general operator. 
So we turn to visualizations of the operator matrix elements
for guidance.
As noted earlier, in practice we do not solve the flow equation
for operators, but apply the unitary transformation 
Eq.~\eqref{eq:Otransform} at $s$ by first
solving the initial and final Hamiltonians for the eigenvectors 
and constructing the matrix
\beqn
 U_{s}(k_{i},k_{j})
   =  \sum_{\alpha}\left<k_i| \psi_{\alpha}(s)\right\rangle\left\langle  
       \psi_{\alpha}(0)|k_j \right>
   \;,
   \label{U:trans}
\eeqn
where $\{k_{i}\}$ is the discrete momentum mesh.

\begin{figure}[tbh]
 \subfigure[]{\includegraphics[width=0.45\textwidth]{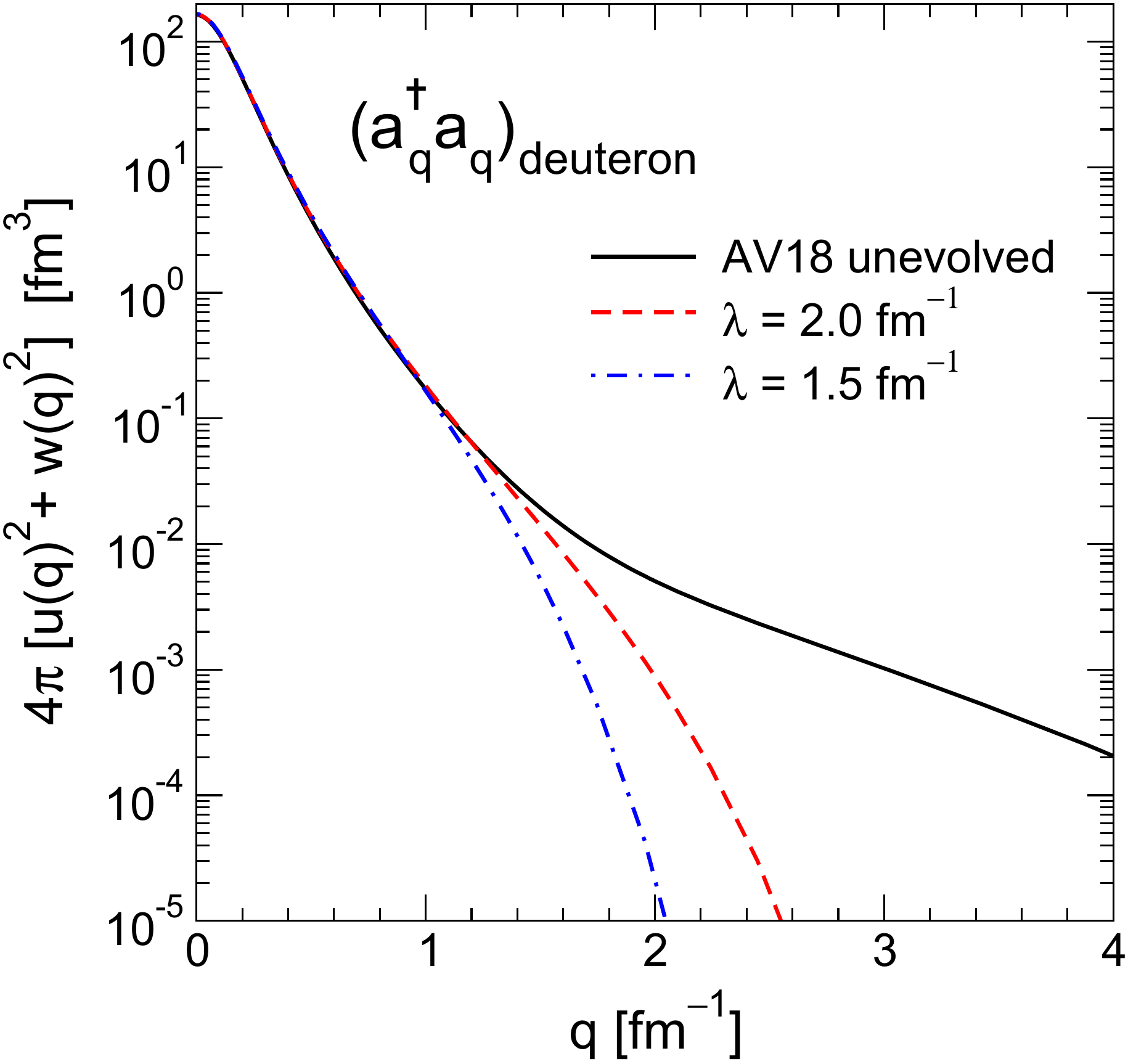}}
 \hspace*{.1in}
\subfigure[]{\includegraphics[width=0.45\textwidth]{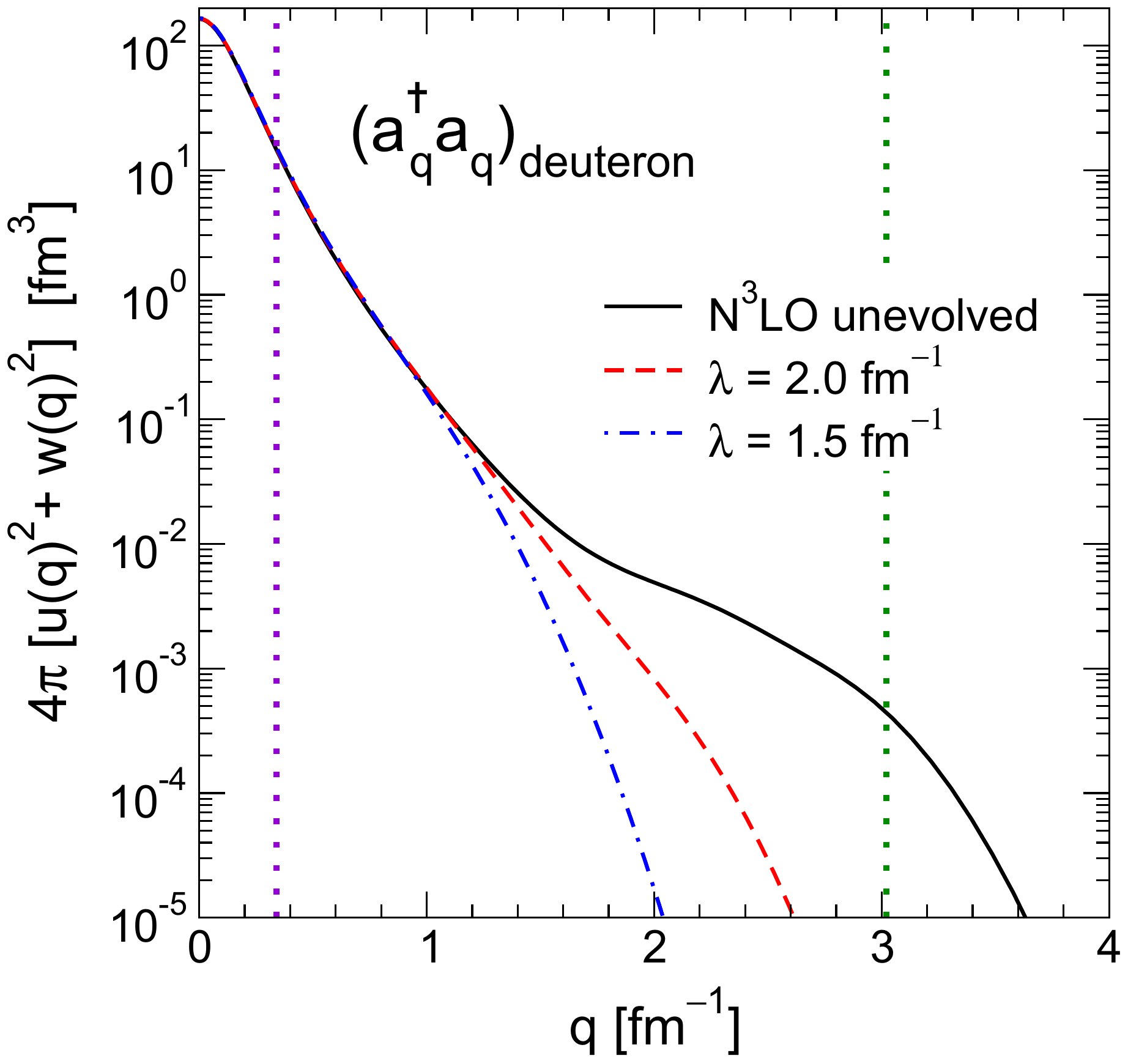}}
\caption{  The momentum distribution in the deuteron 
as given by the expectation value of the
bare number operator $\adaggera$ for the (a) Argonne $v_{18}$ (AV18)~\cite{Wiringa:1994wb} and (b)
N$^3$LO (500\,MeV)~\cite{Entem:2003ft} potentials,
both evolved and unevolved. 
 \ 
 \label{fig:MomDist} }
\end{figure}

\subsection{Momentum Distribution}

We begin our examination of operator evolution with perhaps
the simplest nontrivial example: the momentum occupation operator
$\adaggera$ in the center-of-mass frame (in this frame --- for the A=2 particle space --- \(q\) is the magnitude of both the relative momentum and the single-particle momentum).
By varying the momentum $q$ we can gain insight  
into how the SRG evolution behaves for initial operators dominated
by either high or low momenta.  
In Fig.~\ref{fig:MomDist} the 
plot of the momentum distribution 
is reproduced from the expectation value
$\left\langle\psi_{d} \right 
  \vert a^{\dagger}_{q}a_{q}\left\vert \psi_d \right\rangle$
in the deuteron for the two initial potentials of Fig.~\ref{fig:potential}.  The
solid line is the result when the unevolved wave function (i.e., the
wave function derived from the unevolved potential) is used with the
unevolved \(a^{\dagger}_{q}a_{q}\) operator.  This sets the baseline
for evaluating the effects of the SRG  for each potential.  When one uses the evolved wave function with the evolved
operator, \(U_{s}a^{\dagger}_{q}a_{q}U^{\dagger}_{s}\), the lines are indistinguishable.  
The dashed and dot-dashed lines are calculated using the 
\textit{unevolved} operator
with the \textit{evolved} wave function at \(\lambda=2.0\fmi\) and
\(\lambda=1.5\fmi\), respectively.  These curves quantify
the effect of not consistently evolving  operators and also give 
the momentum dependence of the evolved wave functions.  As can be seen, the high momentum
components of the wave functions are significantly suppressed, as
is consistent with the decoupling seen in the potential. 

\begin{figure}[tbh]
\subfigure[]{\includegraphics[width=0.92\textwidth]{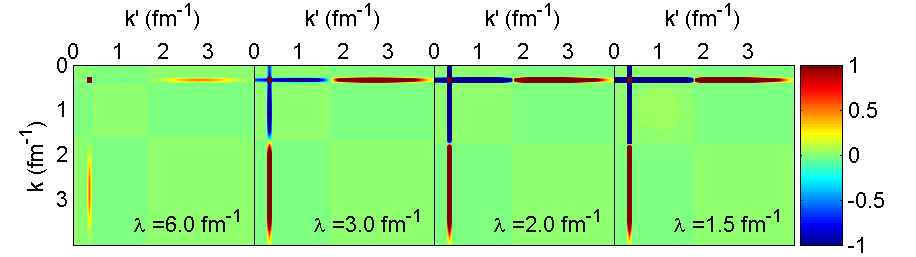}}
{\includegraphics[width=0.92\textwidth]{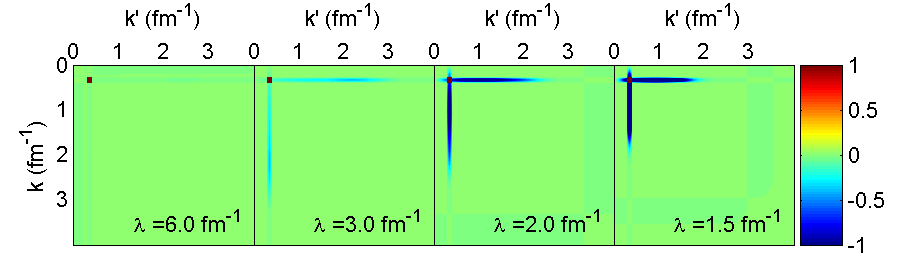}}
\subfigure[]{\includegraphics[width=0.92\textwidth]{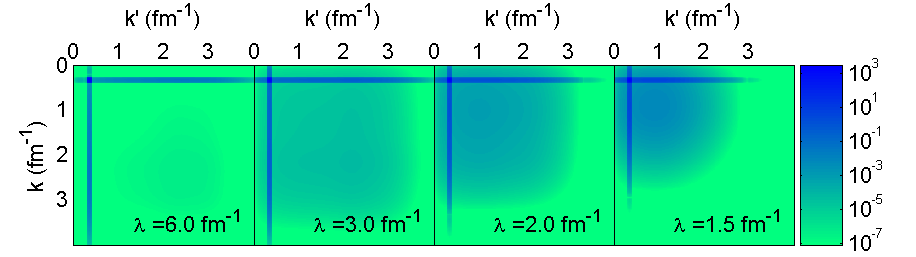}}
\caption{  (Color online) (a) SRG evolution of the operator
\(\left\langle k \right\vert \adaggera\left\vert k' \right\rangle\) for $q=0.34\fmi$ 
in the \(^3\text{S}_1\) partial wave from \(\lambda=6\fmi\) to \(\lambda=1.5\fmi\),
with the N$^3$LO (500\,MeV)~\cite{Entem:2003ft} initial potential.
(b) Integrand of \(\left\langle\psi_{d}(s) \right \vert \adaggeraop \left\vert \psi_d(s) \right\rangle\)  with
 linear (top) and logarithmic magnitude (bottom) scales.}
 \label{fig:NumberIntLow} 
\end{figure}

\begin{figure}[tbh]
\subfigure[]{\includegraphics[width=0.92\textwidth]{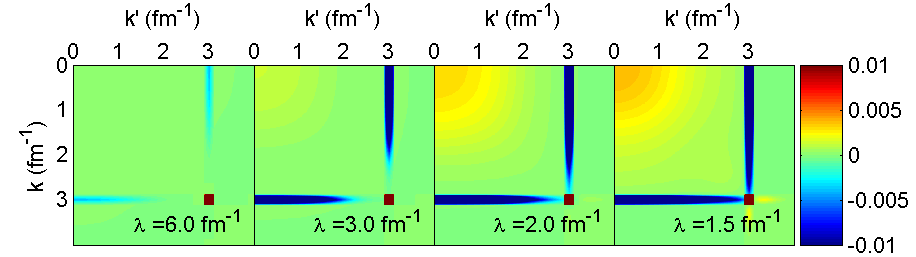}}
{\includegraphics[width=0.92\textwidth]{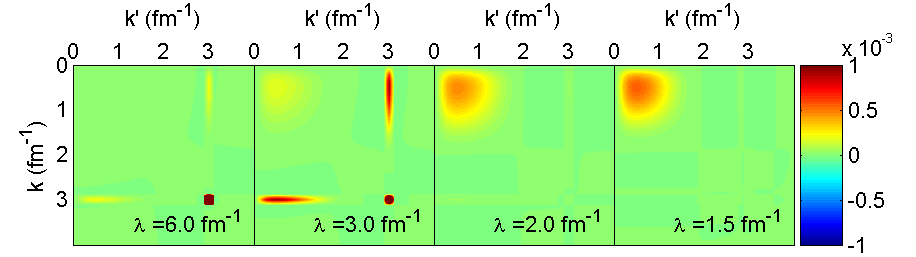}}
\subfigure[]{\includegraphics[width=0.92\textwidth]{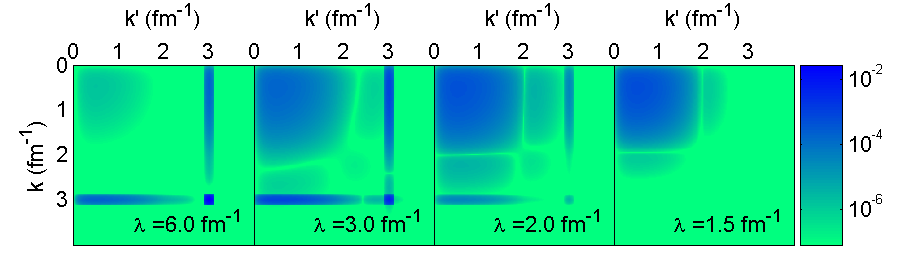}}
\caption{(Color online) Same as Fig.~\ref{fig:NumberIntLow} but for $q=3.02\fmi$.} 
  \label{fig:NumberIntHigh}
\end{figure}

An analogous visual representation 
to that used for the potential
allows us to analyze the RG flow and properties
of  these objects.
Figures~\ref{fig:NumberIntLow} and \ref{fig:NumberIntHigh}
illustrate the flow of the momentum
occupation operator consistent with the renormalization of the N$^3$LO 500 MeV
potential. 
We present the SRG results using only one
potential, but qualitatively similar results will be obtained
using any nuclear potential of interest.  
Each of the figures shows three sequences: the
initial operator matrix elements 
\(\left\langle k \right\vert \adaggera\left\vert k' \right\rangle\)
evolved to four different $\lambda = s^{-1/4}$ values and the integrand
of \(\left\langle\psi_{d}(s) \right \vert \adaggeraop \left\vert \psi_d(s) \right\rangle\) with first linear and then logarithmic scales.
The operators shown
correspond to those used
to calculate the momentum distribution  at \(q=0.34\fmi\) and 
\(q=3.02\fmi\) (marked by the dotted lines
in Fig.~\ref{fig:MomDist}(b)). It is apparent that the unevolved  operator is
simply a delta function in momentum space (some minor evolution can
already be seen at  \(\lambda=6\fmi\) because the scale must
be magnified to view the evolution at lower values of
\(\lambda\)).

\begin{figure}[tbh]
  \includegraphics[width=0.8\textwidth]{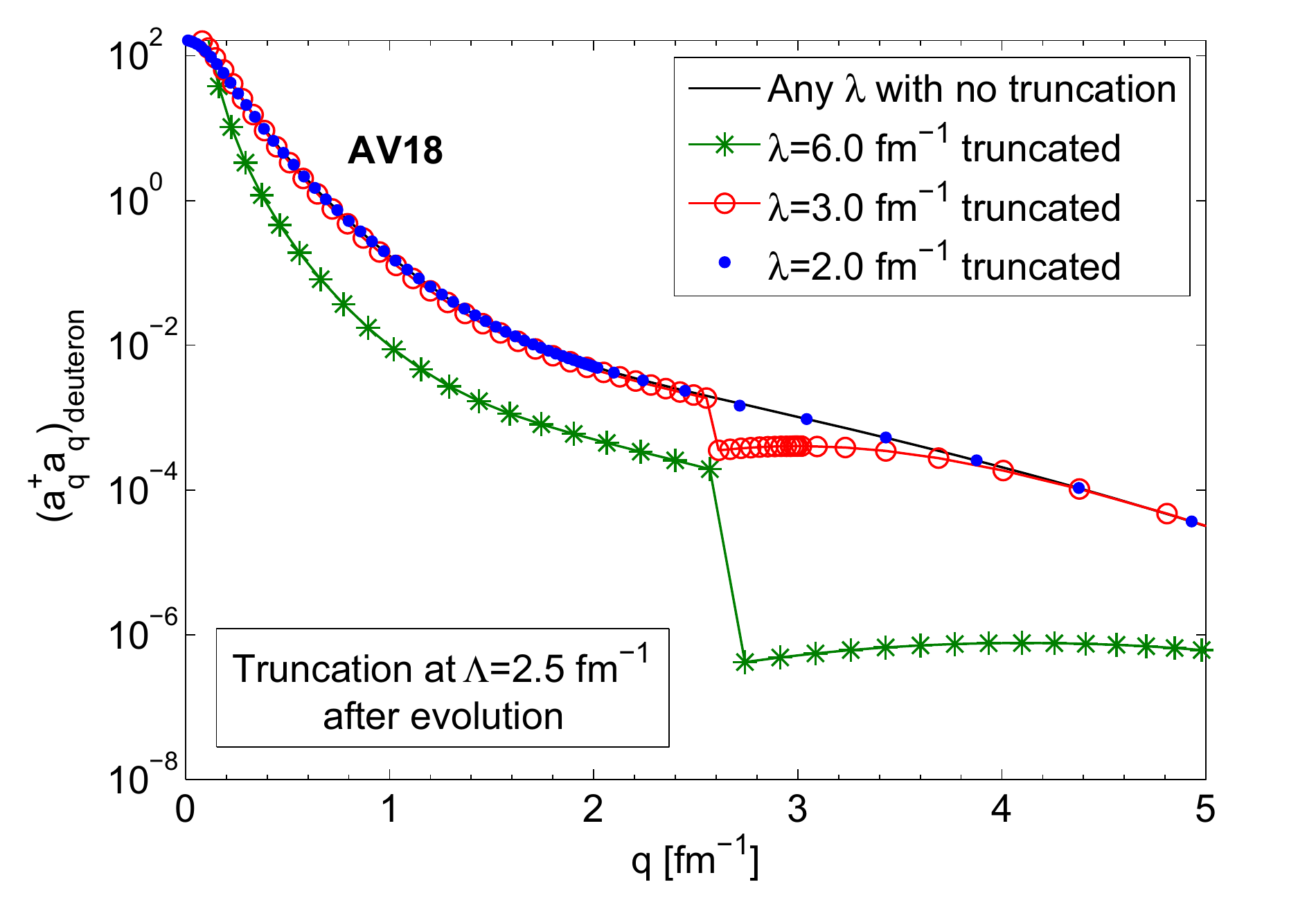}
  \caption{ Decoupling in operator matrix elements is tested
  by calculating the momentum distribution in the deuteron after evolving the AV18
  potential to several different $\lambda$ and then truncating
  the Hamiltonian and evolved occupation operators   (i.e., set them to zero above $\Lambda = 2.5\fmi$).
  }
  \label{fig:MDdecouplingAV18}           
\end{figure}

\begin{figure}[tbh]
  \includegraphics[width=0.8\textwidth]{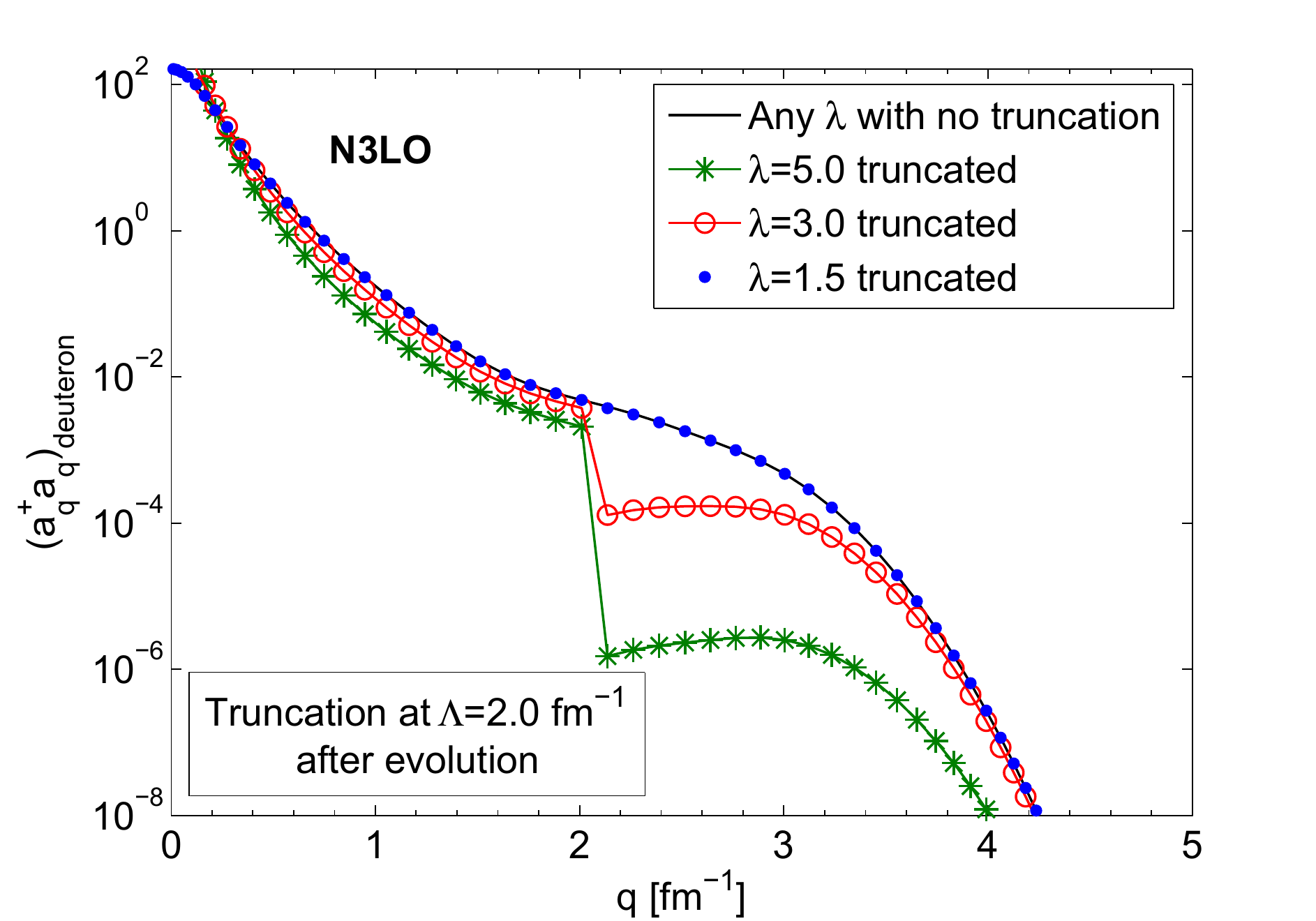}
  \caption{ Decoupling in operator matrix elements is tested
  by calculating the momentum distribution in the deuteron after evolving the N$^3$LO
  potential to several different $\lambda$ and then truncating
    the Hamiltonian and evolved occupation operators   (i.e., set them to zero above $\Lambda = 2.0\fmi$).
  }
  \label{fig:MDdecouplingN3LO}           
\end{figure}

Consider the operator (top row) sequences first.
For both $q$ values, the evolution begins along the momentum axes around
the delta function where we see strength developing that was not
present in the original operator.  This behavior can be understood
from the momentum basis SRG [Eq.~\eqref{O:DEevolve}]
and the  features of the corresponding  potential evolution.
Holding \(k\) fixed, one can see that only the second term in the
integral initially picks up strength along the axis that passes
through the delta function, because the operator is zero everywhere
else, and vice versa holding \(k'\) fixed.  For the operator at high
\(q\) we see that it develops more and more strength at low
momentum as it evolves.   The need for this additional strength is
particularly evident because of the decoupling via evolution of the  
potential and the consequent suppression of high-momentum components
in the deuteron wave function, as seen in Fig.~\ref{fig:MomDist}(b).  As a
result, the operator must pick up additional strength for
the expectation values to remain unchanged. This strength can appear
in two ways: \textit{(i}) It can come in at low momentum, as we see here; or,
\textit{(ii})  the operator can gain strength only at high momenta, where the
operator would have to become pathologically large.  If the second
case were to occur, practical calculations with the SRG in a
reduced basis would not be possible.  
This is found empirically to not be the case, and we can be
confident such pathologies will not occur based on more general
arguments discussed in what follows. 

The operator at low \(q\), however, picks up some strength at larger values of momentum than present in the initial operator.  This is also
needed to compensate for the suppression of low momentum dependence
in high-energy eigenstates to maintain their expectation values.    
One should note that the operator display scale used
here can be a bit deceptive, in that it has been amplified to make the qualitative features of the evolution more apparent.
Most of the evolution does, in fact, remain at low momentum for
deuteron expectation values.

We show the occupation operator as
an integrand given by \(\left\langle\psi_{d} \right \vert
a^{\dagger}_{q}a_{q}\left\vert \psi_d \right\rangle\) in 
Figs.~\ref{fig:NumberIntLow}(b) and \ref{fig:NumberIntHigh}(b).
The expectation value filters the general operator by weighting its matrix elements with
the deuteron wave function. 
Now we can see a clean RG flow in the strength for both operators.  
The integrand of  the operator  at
\(q=3.02\fmi\) begins as a sharp spike, corresponding to the
original operator, but then flows out along the momentum axes to
lower momentum.  By the time the integrand reaches lower values of
\(\lambda\) in the evolution nearly all of the strength in the
expectation value is in the low-momentum region.  The
original spike disappears as the wave function dependence  
at high momentum falls off.  

As for the operator at   \(q=0.34\fmi\), the
strength does begin to flow out to some extent but remains almost
entirely in the low-momentum region.  Once again, the display
scale  has overemphasized the extent of the evolution in the values
of the integrand.  The spike that remains at \(\lambda=1.5 \fmi\)
actually contains \(\approx96\%\) of the full expectation value. 
Owing to the possibility of misinterpreting these plots on a linear
display scale, we also include the same plots with logarithmic
display scale.  
These pictures show
only the magnitude of the integrands, but display nearly the full
range of their values. Now it is conclusive that the strength of the
high momentum operator flows to low momentum, and the strength of
the low-momentum operator remains at low momentum for a low-energy state.  We see this
pattern repeat in the calculation of other operators in the
next section.  

Despite the apparent changes in the
integrands as they evolve, it is important to note that the sum of
all the points (the expectation value) remains unchanged because of the
unitarity of the SRG transformation.
The momentum distribution calculations shown in Fig.~\ref{fig:MomDist} were
performed in the full momentum space of the original
potential.  
Decoupling of the potential
allows us to truncate the model space, thereby making numerical
simulations more feasible, while at the same time allowing us to 
calculate the correct binding energies.  If the calculation of other
expectation values must be performed in the full model space, then
the benefits of the SRG would be lost. 
However, the redistribution of strength implies that we have
a form of decoupling for the operator.
A critical test is to
verify that decoupling is maintained in the calculation of
operator expectation values.  

This check is shown in Figs.~\ref{fig:MDdecouplingAV18} 
and \ref{fig:MDdecouplingN3LO} 
for the momentum distributions of AV18 and N$^3$LO potentials. To
perform the check, we evolved the original Hamiltonians and
operators in the full momentum space to various \(\lambda\) then
truncated the model space to \(\Lambda\).
The deuteron wave
function derived in this truncated space is used to calculate the expectation value of
the evolved number operators to produce the momentum distributions
shown here.  The figures show that when the SRG evolution
\(\lambda>\Lambda\), the curves deviate significantly from that
produced in the full space (because the wave function is
distorted).
However, once the operators are evolved
to \(\lambda\) below the truncation at \(\Lambda\), the expectation
values are reproduced for all values of momenta, even in the region
outside of the new model space.  Thus decoupling is successful
and \(\lambda\)   provides a rough
guide as to where this decoupling occurs. We see that this is
also the case for other operator matrix elements of interest, as
well as understand further how this comes about, in what follows.

\subsection{General Analysis}  \label{subsec:general}

The plots of the deuteron integrands show that no pathologies appear at high momentum in the evolved operators and  
verify that decoupling can be successful when calculating expectation
values of this operator. The plots even indicate where the
model space can be truncated --- this is simply where the integrand
strength becomes negligible in the logarithmic plot.  Here we develop a
more general understanding of operator evolution to build
confidence that
pathologies will not occur in other operators.

 Consider the representation of a
generic operator in terms of the energy eigenstates, 
\beqn
   \widehat{O}_{s} = \sum_{ij}O_{ij}\left\vert 
   \psi_{i}(s)\right\rangle\left\langle  \psi_{j}(s) \right\vert \;, 
\eeqn
where 
\beqn
  O_{ij}= \left\langle  \psi_{i}(s) \right\vert \widehat{O}_{s}\left\vert   
  \psi_{j}(s)\right\rangle
  \;.
\eeqn
It is important to remember that
these matrix elements are invariant under SRG transformations, so \(O_{ij}\) does not depend on \(s\).  Thus,
the momentum space behavior of evolved operators is given, in turn,
by the momentum space behavior of the evolved eigenstates ---
specifically the sum of their outer products weighted by \(O_{ij}\),
\beqn
  O_{s}(k,k')=\sum_{ij}O_{ij}\left<k| \psi_{i}(s)\right\rangle\left\langle  \psi_{j}(s)|k'\right> \;.
\eeqn
The behavior of these eigenstates is well under control.  As we have
seen from the momentum distribution of the deuteron, 
low-energy bound-state
wave functions are suppressed at high momentum.  The rest of the
(positive) eigenstates are effectively smeared out delta
functions (normalized in a finite basis), which because of
decoupling in the Hamiltonian become increasingly
narrow peaks with the evolution in \textit{s}.  
So, the evolution will not become pathological
unless the unevolved operator is already pathological (i.e., only if some \(O_{ij}\) are unnaturally large). 
       
If we now consider this operator in the deuteron eigenstate
$|\psi_d(s)\rangle = |\psi_{1}(s)\rangle$, we find that only the
\(O_{11}\) matrix element of the operator is projected out and the
momentum dependence is given by the outer product of the deuteron
wave function.  Specifically, the only nonzero part of the operator
will be formally given by
\beqn
  O_{s}(k,k') \rightarrow O_{11}
  \left<k| \psi_{1}(s)\right\rangle\left\langle  \psi_{1}(s)|k'\right> \;.
\eeqn  
However, if we would like to reconsider the issue of decoupling and
a finite model space truncation, we find that we are restricted by
the  potential breaking of eigenstate orthogonality in the truncated
space; that is, the extent to which 
\beqn
   \frac{2}{\pi}\int ^{\Lambda}_{0}k^{2}dk\,
     \left\langle  \psi_{i}(s)|k \right\rangle 
     \left\langle k| \psi_{j}(s) \right\rangle \neq0 \;, 
     \text{\quad for}\;i\neq j \;.
\eeqn
The logarithmic integrand plots show us that this problem is
negligible if one of the states is the deuteron.  Furthermore, SRG-driven decoupling of states well separated in energy will make them increasingly orthogonal in the truncated space.  Thus, the  momentum-space evolution of operators in a
specific basis can be brought under control; we will consider the
case of induced many-body components of the operators in a sequel 
to this article~\cite{Anderson:2010aa}.

\section{Other Operators} \label{sec:otherops}

%%%%%%%%%%%%%%%%%%%%%%%%%%%%%%%%%%%%%%%%%%%%%%
\subsection{Long-distance Operators:
  \(\left\langle r^2\right\rangle\), 
\(\left\langle  Q_{d}\right\rangle\), \(\left\langle\oneoverr\right\rangle\)}

We  begin our presentation of additional operators with the
evolution of operators for three paradigmatic expectation values
in the deuteron: the rms radius, the quadrupole moment, and  $\oneoverr$.  
These operators all act on 
relatively long distance scales. At leading order they are
naturally defined in coordinate space so that
deuteron expectation values can be written as  \cite{Bogner:2006vp} %
\beqn
\left\langle r_{d}\right\rangle=\frac{1}{2}\left[\int^{\infty}_{0}\!
  dr\,r^{2}\left(u(r)^{2}+w(r)^{2}\right)
\right]^{1/2}
  \;,
\label{Rsqr:expect}
\eeqn
\beqn
\left\langle Q_{d}\right\rangle=\frac{1}{20}\int^{\infty}_{0}\!
  dr\,r^2 w(r)\left( \sqrt{8}\,u(r)-w(r) \right)
  \;,
\eeqn
and
\beqn
\left\langle\frac{1}{r}\right\rangle=\int^{\infty}_{0}\!dr\,
  \left(\frac{1}{r}\right)\left(u(r)^{2}+w(r)^{2}\right)
 \;,
\label{Rinv:expect}
\eeqn
where \textit{u} and \textit{w} are the \( ^{3}\text{S}_1\) and \(
^{3}\text{D}_1\) deuteron radial wave
functions. However, we  will continue to analyze the operators in
momentum space.  To
avoid numerical instabilities associated with putting the
derivatives in the momentum-space expressions on a mesh
(see Ref.~\cite{Bogner:2006vp}), we
extract the coordinate-space operators from
Eqs.~\eqref{Rsqr:expect}--\eqref{Rinv:expect} and transform to the
partial wave momentum basis.  For example, from
Eq.~(\ref{Rsqr:expect})  we can see that the \(r^{2} \) operator is
given by the diagonal matrix (discretized in coordinate space)
\beqn
\left\langle r \right\vert r^{2}\left\vert r' \right\rangle=r^{2}\delta(r-r')
\eeqn
 in the  \(^{3}\text{S}_{1}\) and \(^{3}\text{D}_{1}\)   channels.
The only transformations needed are given by
\beqn
\label{convert:Coord_Mom1}
\left\langle r|k ;{^{3}\text{S}_{1}}\right\rangle=\sqrt{\frac{2}{\pi}}\,r\,k^{2}\,j_{0}(kr)
\eeqn
and
\beqn
  \left\langle r|k    
   ;{^{3}\text{D}_{1}}\right\rangle=\sqrt{\frac{2}{\pi}}\,r\,k^{2}\,j_{2}(kr)
  \;,
  \label{convert:Coord_Mom2}
\eeqn
where the \(j_{l}\)'s are spherical Bessel functions (additional
partial waves are needed for states other than the deuteron, of
course). The transformations are represented as
\(n\times m\) matrices (where \textit{n} and \textit{m} are the
sizes of the coordinate and momentum space meshes, respectively) and
applied to both sides of the  \(n\times n\)  coordinate-space
operator matrix to produce an \(m\times m\) matrix in momentum space
for each partial wave.  Then, to evolve the operator in momentum
space, we apply the unitary transformation 
\(\Oop_{s}=U_{s}\Oop\,U_{s}^{\dagger}\).
\begin{figure}[tbh]
\includegraphics[width=0.92\textwidth]{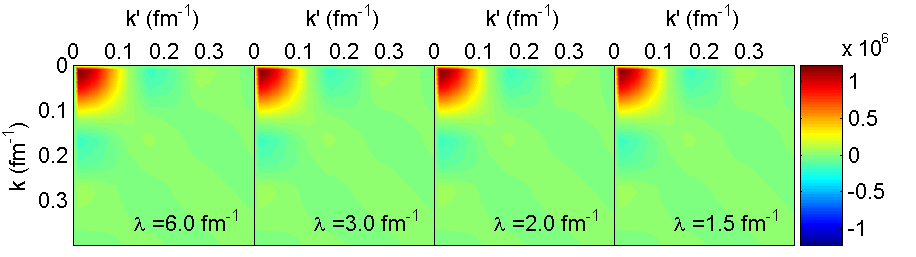}
\caption{(Color online) Operator  evolution of \(\left\langle k \right\vert r^{2}\left\vert k' \right\rangle\) in the  \(^3\text{S}_1\) partial wave from \(\lambda=6\fmi\) to \(\lambda=1.5\fmi\) using the N$^3$LO (500 MeV) potential.  }
\label{fig:RsqrdOp}
\end{figure}

\begin{figure}[tbh]
\includegraphics[width=0.92\textwidth]{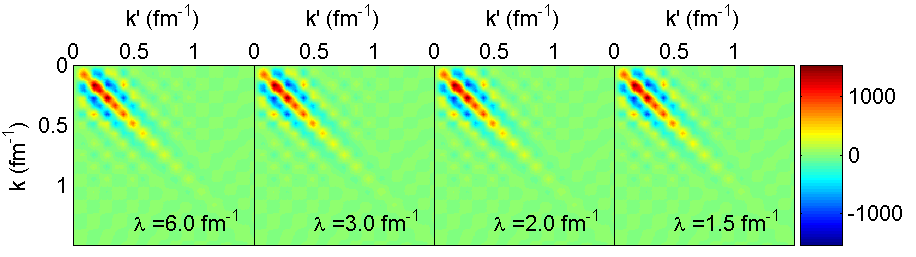}

\includegraphics[width=0.92\textwidth]{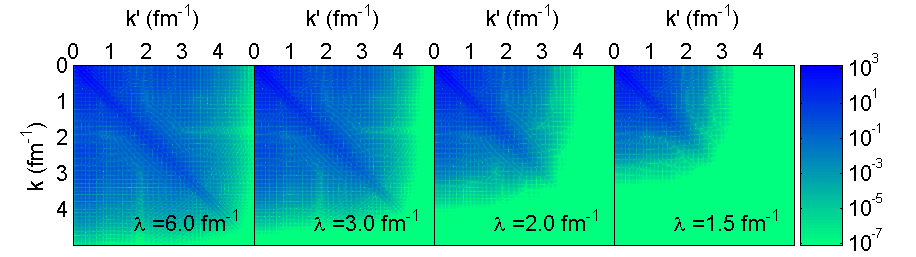}
\caption{(Color online) Integrand given by \(\left\langle\psi_{d} \right \vert r^{2}\left\vert \psi_d \right\rangle\)   and evolved  from \(\lambda=6\fmi\) to \(\lambda=1.5\fmi\) using the N3LO 500 MeV potential with a linear color scale (top) and a logarithmic scale of the magnitude (bottom). Notice the difference in momentum scales. }
\label{fig:Rsqrd}
\end{figure}

The SRG evolution sequence for \(r^{2}\)  shown in Fig.~\ref{fig:RsqrdOp} 
is again a picture of just the operator in the momentum basis.  Note
the restricted momentum scale of the plot, as well as the magnitude of the
operator display scale.  As a long distance operator,  the strength
is highly concentrated at low momentum.  It is evident that very little renormalization occurs at the lowest
values of momentum, which is consistent with the findings for the
number operator.  Looking at the deuteron integrand of the \(r^{2}\) operator
in Fig.~\ref{fig:Rsqrd}, the linear display scale plot shows the
same behavior.  The log-scale
shows the entirety of the
contributions to the integral.  At low momentum, the
strength is orders of magnitude greater than 
elsewhere.  While the pattern of the contribution
changes very little with \(\lambda\) in the linearly scaled
plot, the log-scale plot clearly shows the characteristic exponential
suppression with decreasing $\lambda$
due to the decoupling of the potential and consequent
reduction of high-momentum components in the deuteron wave function.

\begin{figure}[tbh]
\includegraphics[width=0.92\textwidth]{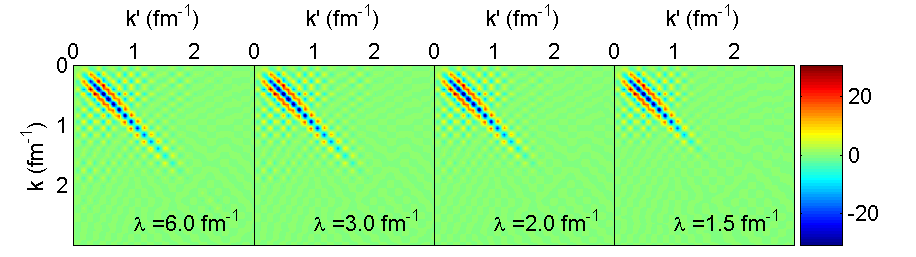}

\includegraphics[width=0.92\textwidth]{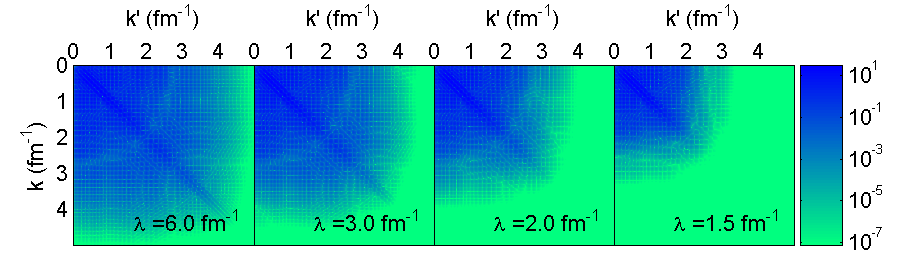}
\caption{(Color online) Integrand of the quadrupole moment expectation value in the deuteron and evolved  from \(\lambda=6\fmi\) to \(\lambda=1.5\fmi\) using the N3LO 500 MeV potential with a linear color scale (top) and a logarithmic scale of the magnitude (bottom). Notice the difference in momentum scales. }
\label{fig:Quadrupole}
\end{figure}

\begin{table}[t]%%h is inline??
\begin{tabular}{|c|c|c|c|c|}\hline
$\lambda$ & $6.0\fmi$ & { \(3.0\fmi\) }& { \(2.0\fmi\) }& { \(1.5\fmi\)} 
  \\ \hline
 $Q_{d}$ & $ 0.275 \fm^2$ & { \( 0.274 \fm^2\)} & { \( 0.269 \fm^2\)} & { \( 0.260 \fm^2\)} \\\hline
\end{tabular}
\caption{Expectation value of the quadrupole moment given by the unevolved operator with the evolved deuteron wave function.  }
\label{table:quadrupoleBare}
\end{table} 

The quadrupole moment operator is also a long-distance operator
and we find its 
evolution shares most of the
same characteristics found for the evolution of the  \(r^{2}\)
operator (note that it  also picks up
strength in the $^{3}\text{S}_{1}$--${^{3}\text{D}_{1}}$ channel).
The deuteron integrand  is shown in Fig.~\ref{fig:Quadrupole}. There
is little  change in the actual evolution of the
operator, while high-momentum contributions become exponentially
suppressed.  
The lack of evolution can be quantified by calculating the
expectation value of the unevolved operator with the evolved
wave function.  Results for various $\lambda$ are given in
Table~\ref{table:quadrupoleBare}. 
Note that we are using the basic, one-body quadrupole moment
operator without two-body or other higher order renormalized
corrections.  The ``true" value for this potential and operator is
  \(Q_{d}\approx 0.275 \fm^2\) while the experimental
value of this quantity is \(Q_{expt.}\approx
0.285\fm^2\).  Thus the induced two-body 
contribution is the same order as omitted two-body contributions
to the initial operator.
As such, the SRG has not led to
any changes larger than one would expect from including the fully
renormalized operator from the EFT.  
  
\begin{figure}[tbh]
\includegraphics[width=0.92\textwidth]{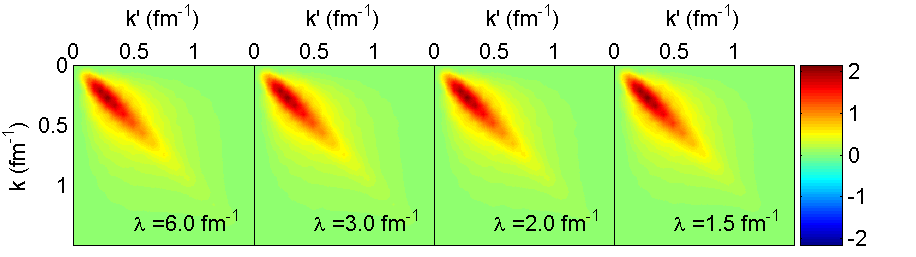}

\includegraphics[width=0.92\textwidth]{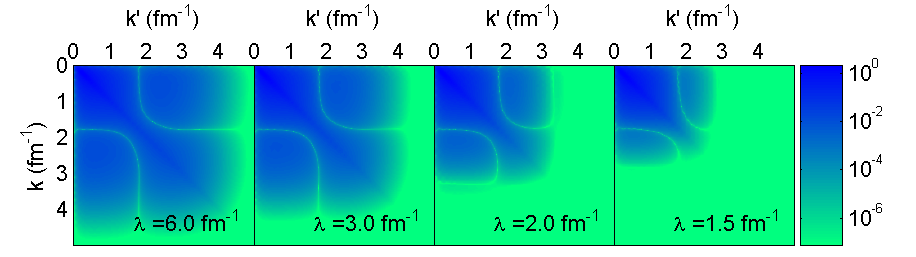}
\caption{(Color online) Integrand given by \(\left\langle\psi_{d} \right \vert \oneoverr\left\vert \psi_d \right\rangle\)   and evolved  from \(\lambda=6\fmi\) to \(\lambda=1.5\fmi\) using the N3LO 500 MeV potential with a linear color scale (top) and a logarithmic  scale of the magnitude (bottom). Notice the difference in momentum scales.  }
\label{fig:Rinv}
\end{figure}

Finally,
in Fig.~\ref{fig:Rinv} we show the evolution in the deuteron of the
$\oneoverr$ operator,  which has larger contributions at short
range than the previous examples. Consequently, this operator is
more spread out in momentum space.  Yet, we see the same general behavior
with respect to renormalization and the suppression 
at large momenta without, as usual, any changes in the expectation
value. 

%%%%%%%%%%%%%%%%%%%%%%%%%%%%%%%%%%%%%%
\subsection{Deuteron form factors: \(G_{\rm C}\), \(G_{\rm Q}\), and \(G_{\rm M}\). }

We now turn to the SRG evolution of electromagnetic  operators that
determine the deuteron charge, quadrupole, and magnetic form factors (i.e., 
\(G_{\rm C}\), \(G_{\rm Q}\), and \(G_{\rm M}\) respectively)~\cite{Sick:2001rh,Garcon:2001sz,Gilman:2001yh,Phillips:2006im,Valderrama:2007ja}.
We restrict our discussion to
deuteron expectation values that have been derived consistently with chiral \(\text{EFT}\)  at
leading order in coordinate space \cite{Valderrama:2007ja}.  These
are given by
\beqn
  G_{\rm C}(Q^{2})=G_{\rm E}^{(s)}(Q^{2})\int \!  dr\,[u^{2}(r)+w^{2}(r)] 
    \,j_{0}\left(|\boldsymbol q|r/2\right)
  \;,
\eeqn
\beqn
  G_{\rm Q}(Q^{2}) = G_{\rm E}^{(s)}(Q^{2}) \frac{6\sqrt{2}}{Q^{2}}
    \int\! dr\, \left[u(r)w(r)+ {w^{2}(r)}/{\sqrt{8}}\right]
      \, j_{2}\left({|\boldsymbol q|r}/{2}\right)
  \;,
\eeqn
\begin{eqnarray}
   G_{\rm M}(Q^{2}) &=& 
      G_{\rm E}^{(s)}(Q^{2}) \frac{3}{2}\int\! dr\,
       w^{2}(r)\left[j_{0}\left({|\boldsymbol q|r}/{2}\right)
      +j_{2}\left({|\boldsymbol q|r}/{2}\right)\right]
      \nonumber \\ & & \quad \null
      + G_{\rm M}^{(s)}(Q^{2})2 \int\!  dr\,
      u^{2}(r)\,j_{0}\left({|\boldsymbol q|r}/{2}\right)
      \nonumber \\ & & \quad \null
      + G_{\rm M}^{(s)}(Q^{2}) \{ \sqrt{2}\int\! dr\,
      u(r)w(r)\,j_{2}\left({|\boldsymbol q|r}/{2}\right)
      \nonumber \\ & & \qquad\qquad\quad \null
      - \int\! dr\, w^{2}(r)\left[j_{0}\left({|\boldsymbol q|r}/{2}\right)
      -j_{2}\left({|\boldsymbol q|r}/{2}\right)\right] \}
      \;,
\end{eqnarray}
where  \(Q^{2}=|\boldsymbol q|^{2}\), and \(G_{\rm E}^{(s)}(Q^{2})\)  and
\(G_{\rm M}^{(s)}(Q^{2})\) are the single-nucleon
isoscalar electric and magnetic form
factors obtained from the parametrization given in
Ref.~\cite{Belushkin:2006qa}. From these coordinate-space expressions, 
we can apply the same procedure used earlier to extract
the operators, then use
Eqs.~\eqref{convert:Coord_Mom1} and \eqref{convert:Coord_Mom2} to convert
to momentum space and transform via the SRG unitary transformation.  One
should  note that starting from a coordinate-space operator is by no
means essential; it is simply a numerical convenience in this case. 
\begin{figure}[tbh]
\subfigure[]{\includegraphics[width=0.48\textwidth]{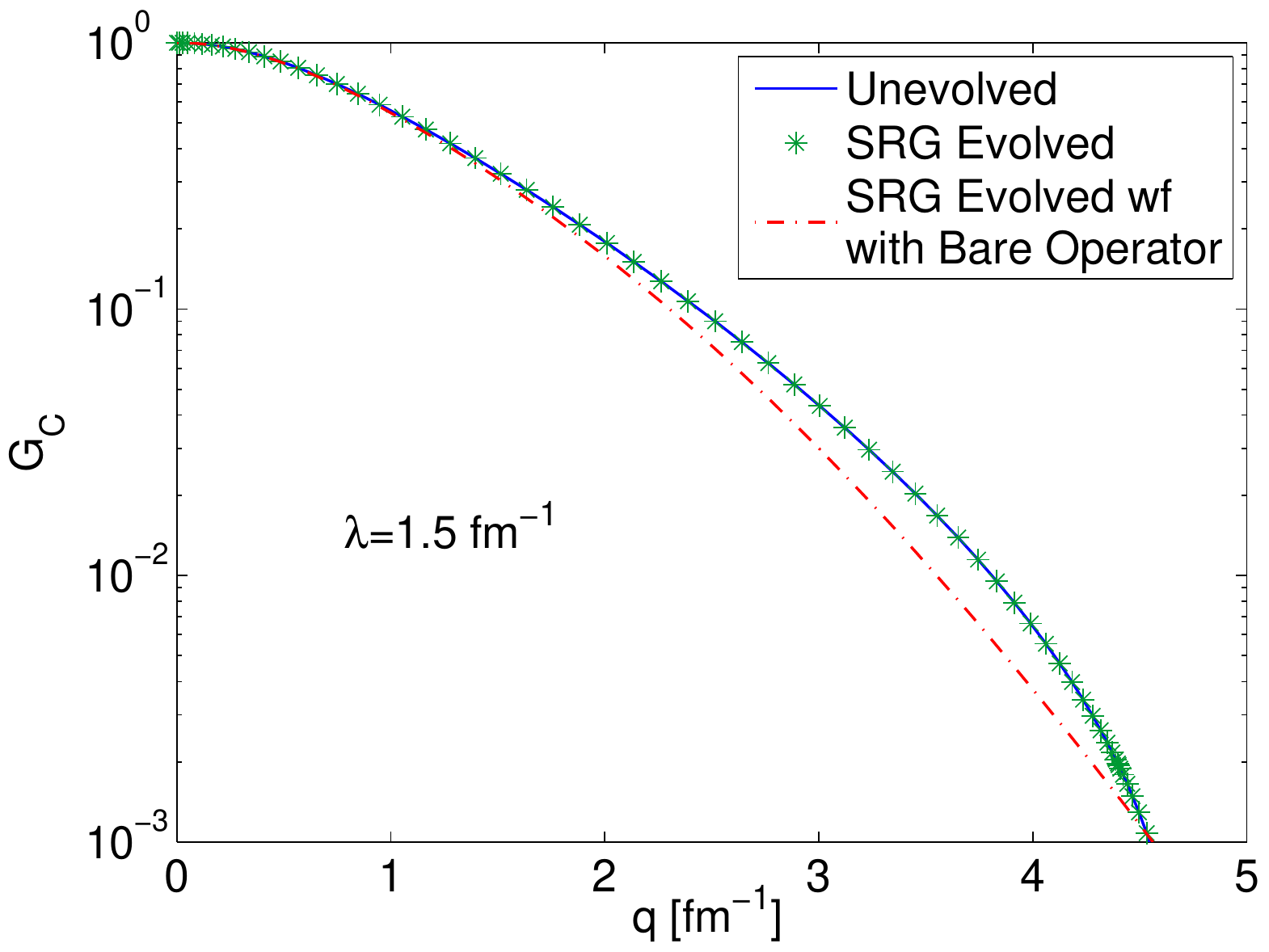}}
\subfigure[]{\includegraphics[width=0.48\textwidth]{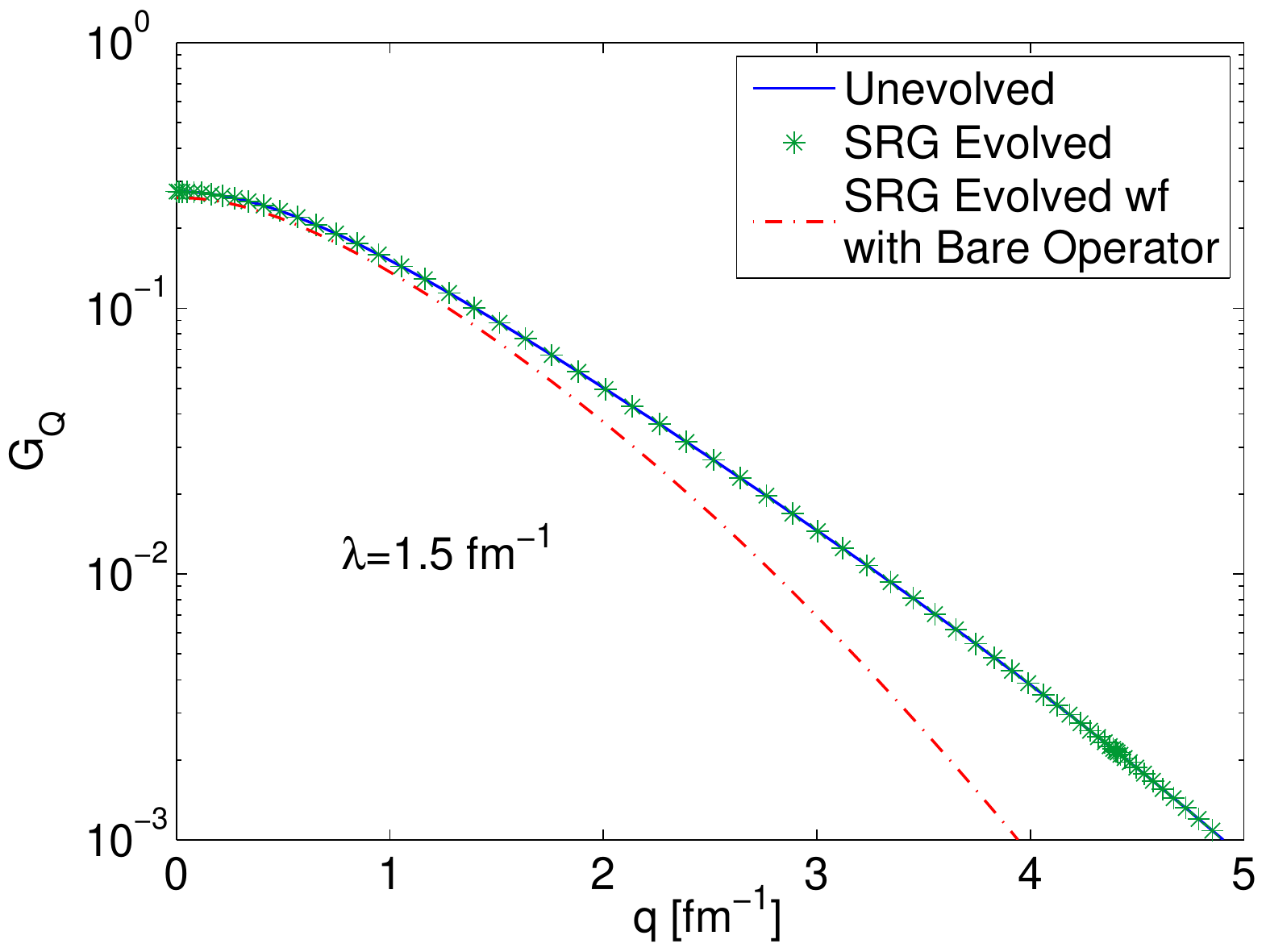}}
\subfigure[]{\includegraphics[width=0.48\textwidth]{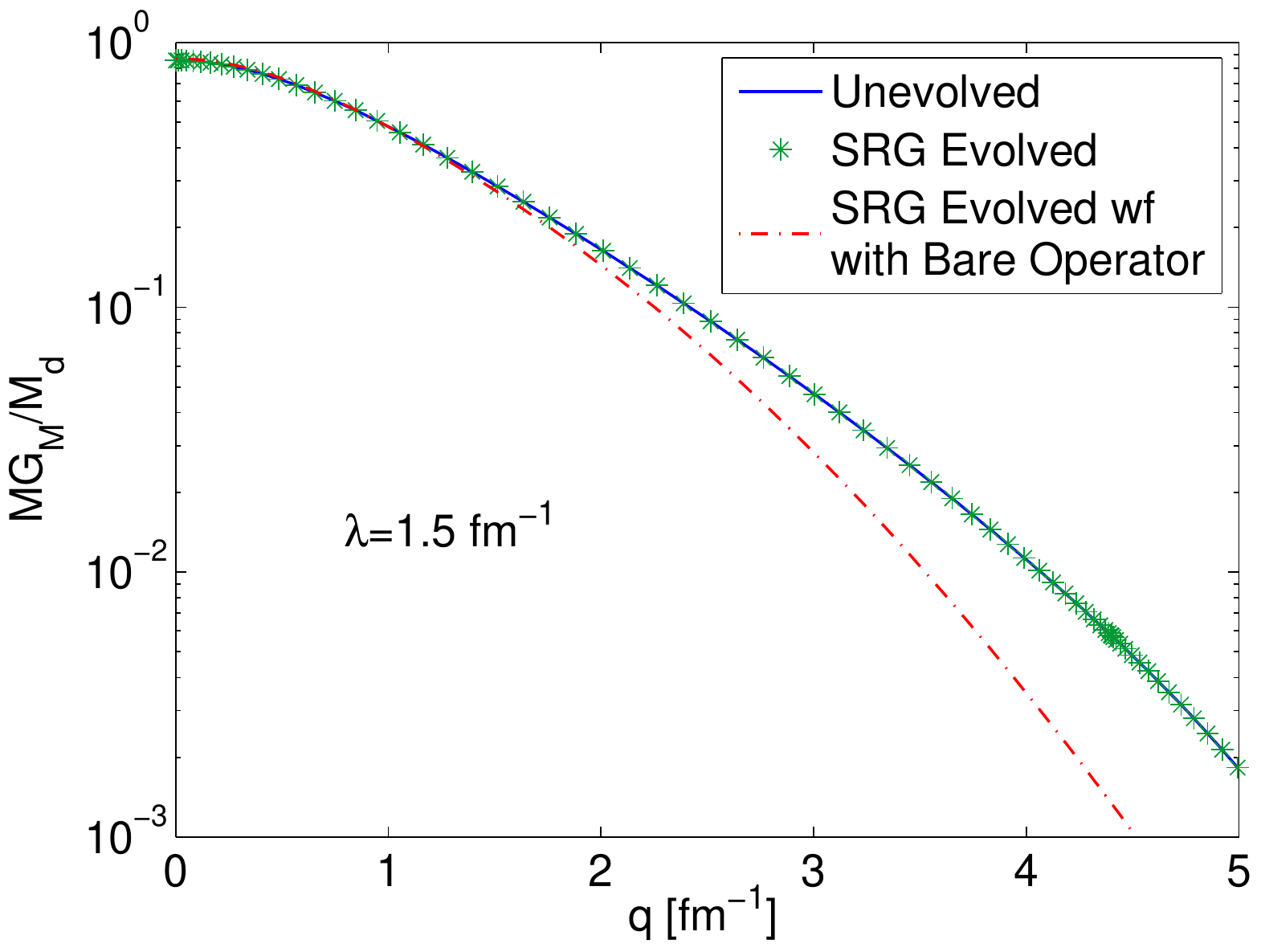}}
\caption{Deuteron form factors   \(G_{\rm C}\), \(G_{\rm Q}\), and \(G_{\rm M}\)  using the isoscalar electric form factor parametrization from Ref.  \cite{Belushkin:2006qa}. \(M\) is the  nucleon mass and \(M_{d}  \) is the mass of the deuteron.  The wave function is derived from the NNLO 550/600 MeV potential and the evolution is run to \(\lambda=1.5\fmi\) \cite{Epelbaum:2004fk}.}
\label{fig:FormFactors}
\end{figure}

The expectation values as functions of $q = \sqrt{Q^2}$ for each of these
operators is presented in Fig.~\ref{fig:FormFactors}.  The solid
line has been calculated using the unevolved potential with the
unevolved operator, which again serves as our reference value. 
The starred points are calculated using the evolved wave function with
the evolved operator (both at \(\lambda=1.5\fmi\)).  As advertised,
they lie precisely on top of the solid line for all values of
$q$, up to small numerical errors.  The  dot-dashed line
is calculated using the \textit{unevolved} operator with
the \textit{evolved} wave function as an indication of the effect of
renormalization on the expectation value.  We see
noticable deviation above \(q\sim\lambda\); however, from the magnitude
of the suppression seen in the wave function at high momentum one
might have expected the curve to drop much faster with respect to
$q$.  However, the form factor
operators probe momenta in the deuteron center-of-mass frame whereas
$q$ is specified in the laboratory frame.  Thus, the operators
are probing the wave function largely at \(\frac{1}{2}q\),  which
again brings the calculations in line with our SRG
expectations. 

\begin{figure}[tbh]
\includegraphics[width=0.92\textwidth]{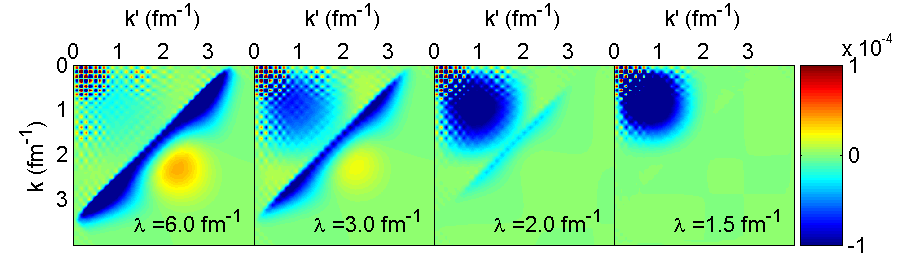}

\includegraphics[width=0.92\textwidth]{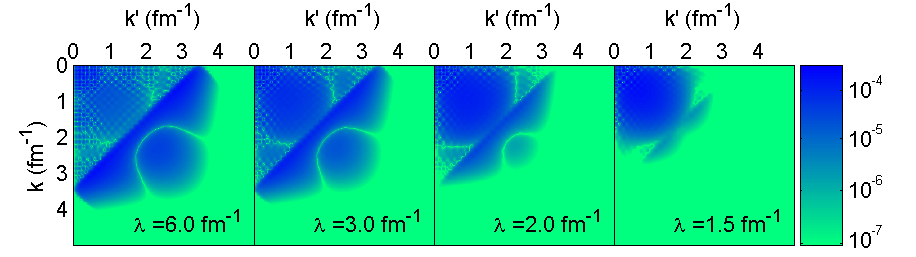}

\caption{(Color online) Integrand of \(G_{\rm M}\) at  \(q=6.90\fmi\)
evolved  from \(\lambda=6\fmi\) to \(\lambda=1.5\fmi\) using the
NNLO 550/600 MeV potential with a linear scale (top) and a
logarithmic scale of the magnitude (bottom).    }
\label{fig:GmHigh}
\end{figure}

\begin{figure}[tbh]
\includegraphics[width=0.92\textwidth]{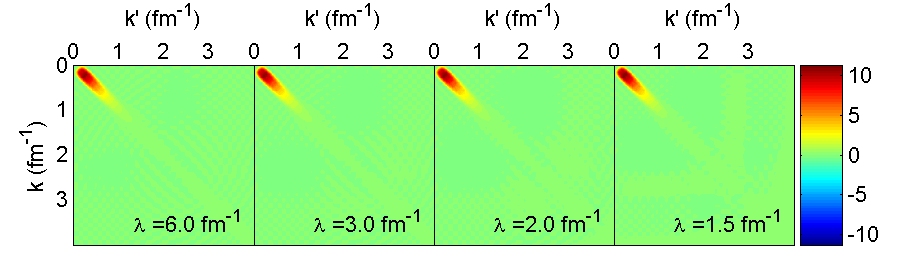}

\includegraphics[width=0.92\textwidth]{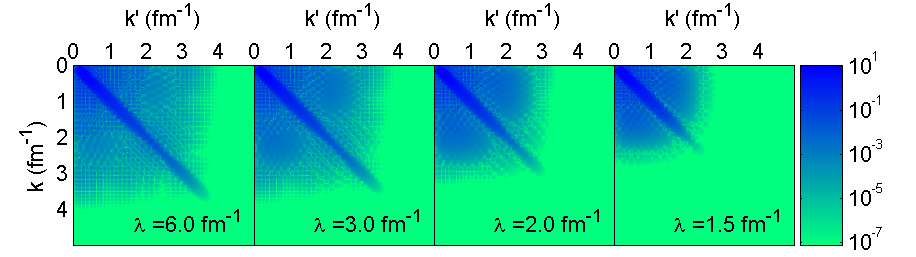}
\caption{(Color online) Integrand of \(G_{\rm M}\) at  \(q=0.34\fmi\)
evolved  from \(\lambda=6\fmi\) to \(\lambda=1.5\fmi\) using the
NNLO 550/600 MeV potential with a linear scale (top) and a
logarithmic scale of the magnitude (bottom).  }
\label{fig:GmLow}
\end{figure}

The basic features of the SRG evolution of all three operators are
qualitatively very similar, so we present the visual matrix representation of the
magnetic form factor at high and low \textit{q} as a representative 
example in Figs.~\ref{fig:GmHigh}   and
\ref{fig:GmLow}. This form factor picks up
strength in all deuteron channels. The high-momentum form
factor has a much greater diffusion of strength at high momentum
than seen in any of the static properties explored earlier.  Yet it
is apparent that the strength in the operator flows to low momentum
in this case also.  In contrast, the low-momentum operator exhibits
very little renormalization, as we have come to expect.  For both
cases, we can see from the  logarithmic display scale plots that the
momentum dependence of the  form factors  is virtually eliminated at
large momenta without affecting the outcome of the computation.

This lack of dependence on high momenta in evaluating 
expectation values is particularly significant for the practical
application of the SRG in calculations of low-energy few- or many-body
systems.     Had this renormalization led to singular behavior in
the operators at high momentum, the effects due to the suppression
of the wave function would have been negated and led to wildly
erroneous results in the evaluation of observables in a reduced
model space. The arguments in Sec.~\ref{subsec:general} 
explain why this will generally be the case.

%%%%%%%%%%%%%%%%%%%%%%%%%%%%%%%%%%%%%%%%%%%%%%%%%%%
\section{Variational Calculations}  \label{sec:variational}

\begin{figure}[tbh]
  \includegraphics[width=0.67\textwidth]{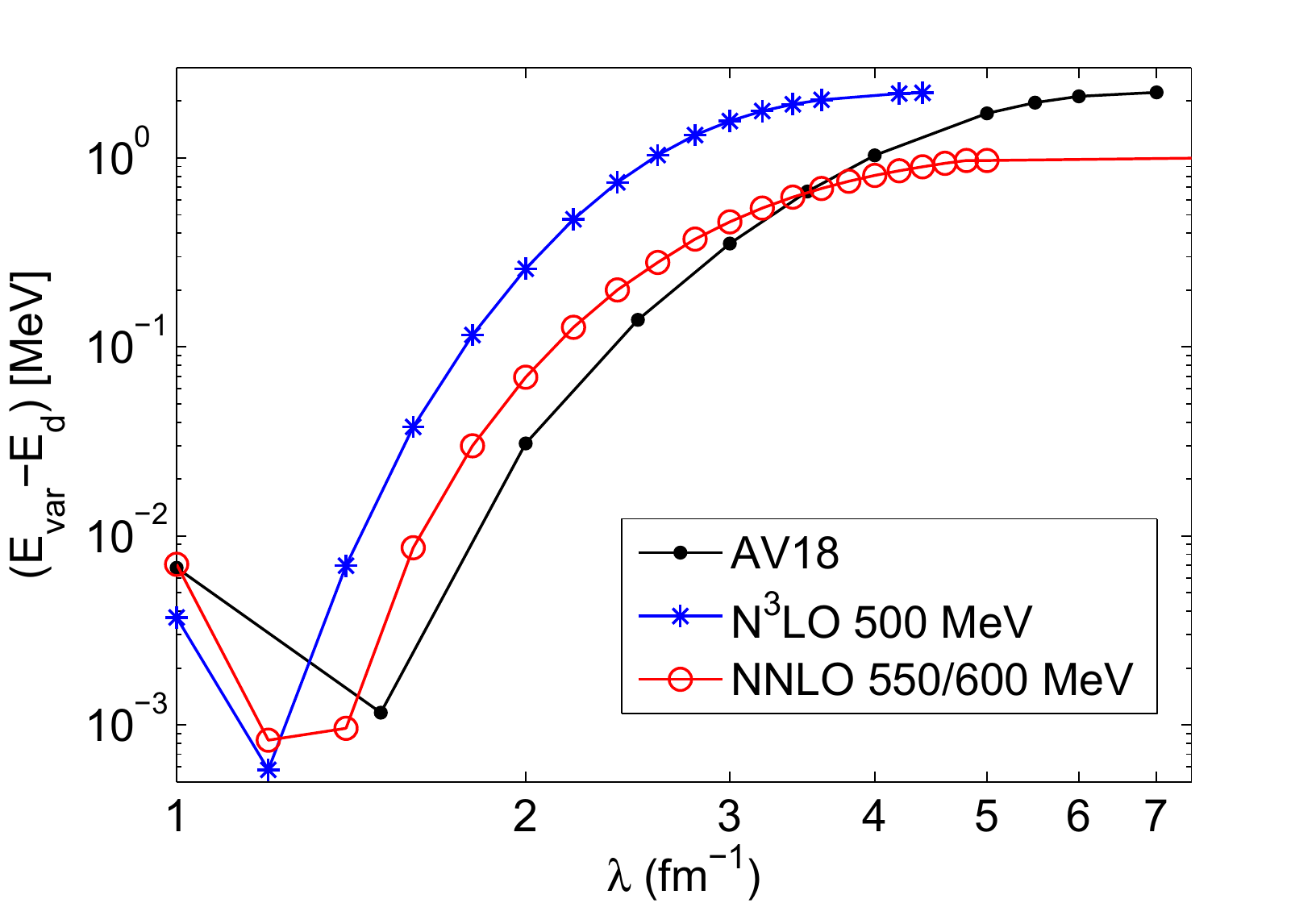}
  \caption{ Deviation from \(E_{d}\) of the best variational energy as a function of SRG\ decoupling parameter for the wave function ansatz of Eq. (\ref{eqn:VariationalAnsatz}).  \(E_{d}\) is the deuteron binding energy for each interaction derived via a full eigenvalue calculation of the Hamiltonian.  
  }
  \label{fig:VariationalEnergies}           
\end{figure}

\begin{figure}[tbh]
  \includegraphics[width=0.67\textwidth]{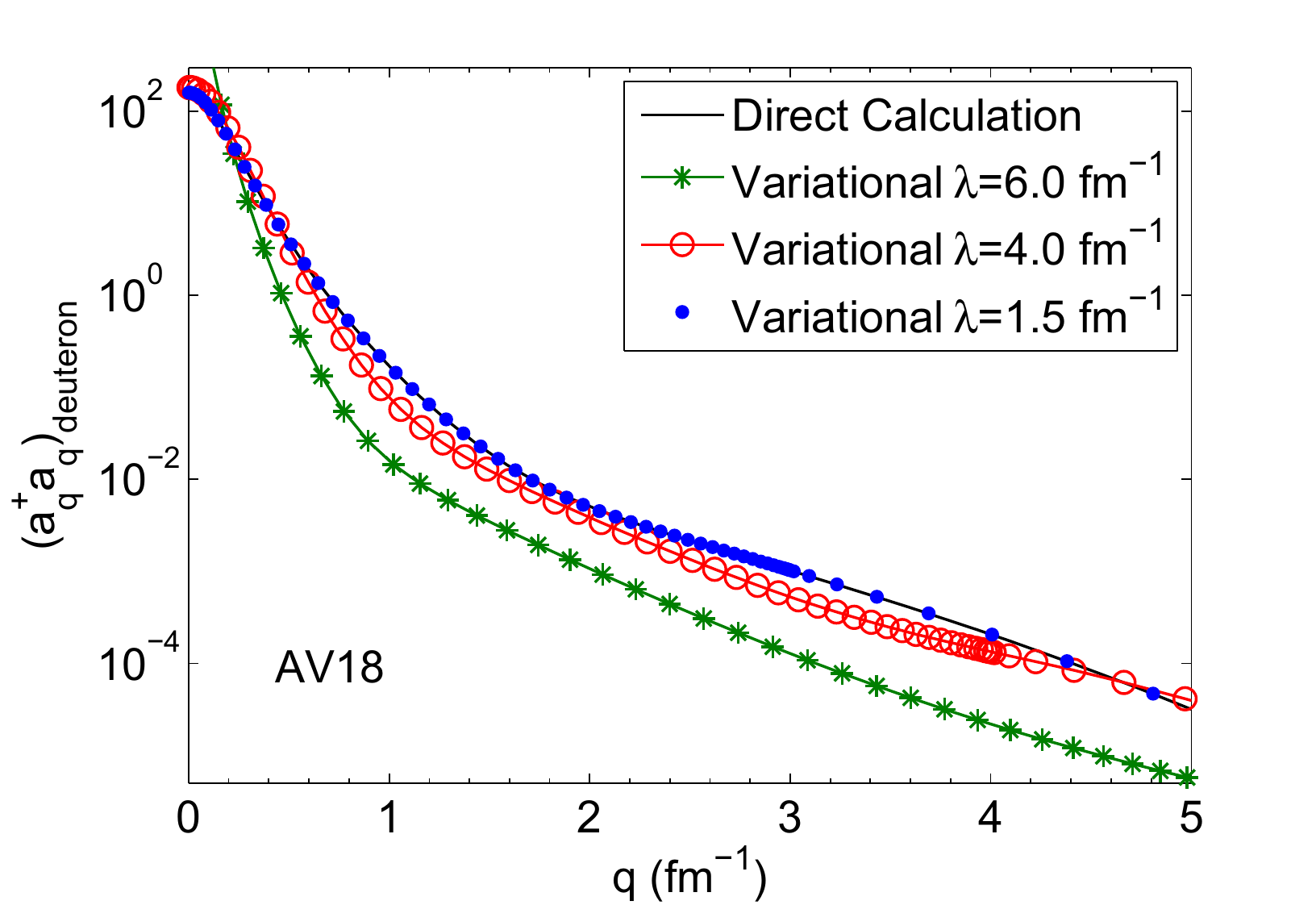}
  \caption{ The momentum distribution in the deuteron 
as given by the expectation value of the
evolved occupation operator $U\adaggera U^{\dag}$ 
using the variational wave functions derived from the Salpeter  ansatz and the AV18~potential  evolved to \(\lambda=6.0\fmi, ~4.0\fmi  \) and \(1.5\fmi\).  The direct calculation is from a full eigenvector solution of the Hamiltonian. 
  }
  \label{fig:VariationalNumOp}           
\end{figure}

\begin{figure}[tbh]
  \includegraphics[width=0.66\textwidth]{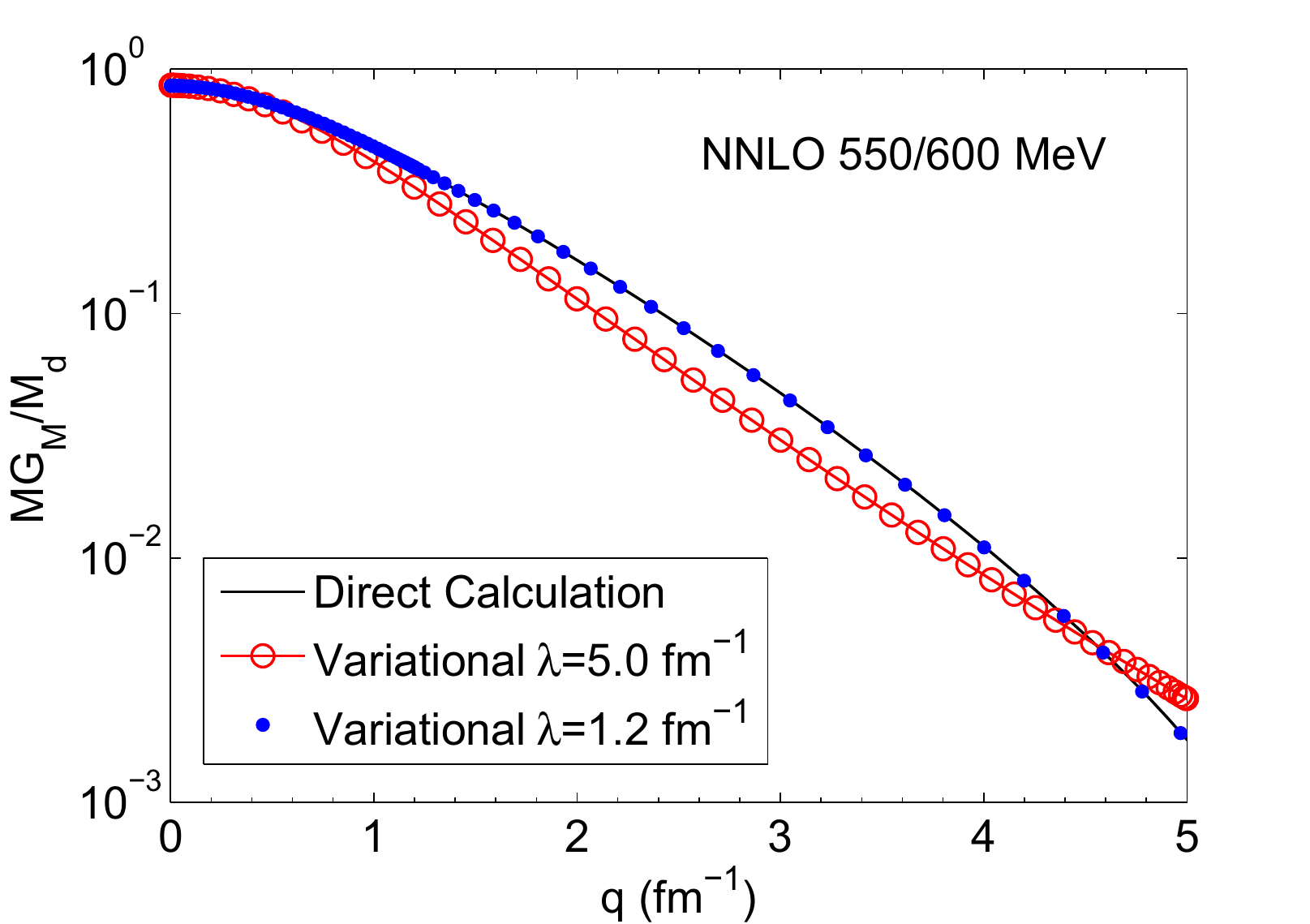}
  \caption{ Deuteron form factor \(G_{\rm M}\)  using the isoscalar electric form factor parametrization from Ref.   \cite{Belushkin:2006qa}. \(M\) is the  nucleon mass and \(M_{d}  \) is the mass of the deuteron.  The variational wave functions are  derived from the Salpeter  ansatz and the NNLO 550/600 MeV potential  evolved to \(\lambda=5.0\fmi\) and \(\lambda=1.2\fmi\). The operators are evolved consistently.  The direct calculation is from a full eigenvector solution of the Hamiltonian. 
  }
  \label{fig:VariationalGMFormFactor}           
\end{figure}

We have argued previously that SRG-evolved interactions and the resulting wave functions become ``simpler."  Variational calculations of the ground-state wave function  can provide a test (and  additional meaning) of the extent to which this is true in practical applications. The decoupling of high- and low-momentum states caused by SRG evolution means that the resulting wave functions are much less correlated than the original wave functions.  Consequently, one expects  that  variational calculations of the evolved wave functions  should be effective with a much simpler ansatz  than one would normally require for the corresponding unevolved interaction.    In conjunction with this, one might be concerned that a delicate interplay of the  evolved operators and wave functions  would be necessary to preserve matrix elements.   However, it turns out  that the evolved operators are not only equally good, but actually superior to the original operators in this respect.
Rather than high-momentum operators picking up small
pieces of the wave function (which could never be reproduced by a
simple variational calculation), we get a smooth sum over where the
wave function is large and easily approximated.  We illustrate these points here  by choosing a simple variational ansatz and looking at how convergence improves for SRG-evolved operator expectation values.    

Variational calculations have been performed on low-momentum potentials (specifically ``$\vlowk$," derived using an alternative RG formulation) in the past to demonstrate a significantly improved convergence  of the binding energies \cite{Bogner:2005fn,Bogner:2006ai}. We choose to adapt a simple ansatz for the deuteron used in those calculations to make our point for the SRG.  In particular, we take the (unnormalized) \(^{3}S_{1}\) and \(^{3}D_{1}\) partial waves to be 
\begin{eqnarray}
  u(k)&=&\frac{1}{(k^{2}+\gamma^{2})(k^{2}+\mu^{2})}e^{-\left( \frac{k^{2}}{\lambda^{2}} \right)^{2}},
       \\ 
    w(k)&=&\frac{ak^{2}}{(k^{2}+\gamma^{2})(k^{2}+\nu^{2})^{2}}e^{-\left( \frac{k^{2}}{\lambda^{2}} \right)^{2}},
\label{eqn:VariationalAnsatz}
\end{eqnarray}
where \(\gamma,\mu,\nu,\) and \(a\) are variational parameters.
The exponential factors are chosen to match the asymptotic suppression of the wave function
resulting from the decoupling of the interaction according to Eq.~\eqref{eqn:ApproxDecoupling}.
  The energy is minimized with respect to the variational parameters at various \(\lambda\) for the three different potentials used in this article.  The binding-energy results are shown in  Fig. \ref{fig:VariationalEnergies}.  Without evolution, the AV18 and N\(^{3}\)LO trial wave functions are not even bound, and the NNLO wave function  accounts for less than half the binding energy.  With evolution, the AV18 and N3LO wave functions begin to bind at \(\lambda\approx7\fmi\) and \(\lambda\approx4.5\fmi\) respectively and when the evolution is taken further, the trial wave function is able to reproduce the exact binding energies to within \(\approx1\,\text{keV}\).

Examining the matrix elements of operators which initially have strength concentrated over a range of different momenta --- such as the occupation operators with respect to momenta \(q\) --- provides a stricter test of the variational solution to the evolved wave functions and the sensitivity of evolved operators to them.  The initial AV18 potential has particularly strong correlations at high momenta.  If we look at the evolved occupation operators in Fig. \ref{fig:VariationalNumOp}, we see three curves: one with a wave function near the binding threshold, one at about half the binding energy, and one that is well converged with respect to the binding energy (evolved to \(\lambda=6.0\fmi, ~4.0\fmi  \) and \(1.5\fmi\) respectively). Near the binding threshold, the momentum distribution is reproduced rather poorly, but at smaller \(\lambda\) the curve improves, and once the binding energies are converged the operator expectation values are also converged to approximately 1\% or better. 

The same pattern holds for other operators and interactions; that is, the operator matrix elements are not sensitive to the fine details of the evolved wave function.  The magnetic form factor of the deuteron using the NNLO potential,  for example, is shown in Fig. \ref{fig:VariationalGMFormFactor}.   Not only does this operator pick up strength in both partial wave states of the deuteron, but also their coupling.  Again, for a variational wave function at  half binding energy (\(\lambda=5.0\fmi\)) the matrix elements deviate significantly from the direct, non-variational calculation, but when the binding energy is converged (at \(\lambda=1.2\fmi\)), the form factor expectation values are reproduced to better than 1\%.

%%%%%%%%%%%%%%%%%%%%%%%%%%%%%%%%%%%%%%%%%%%%%%%%%%%
\section{Operator Factorization}  \label{sec:opfactor}

In this section, we consider in more detail the expectation value 
in a low-energy bound state of
operators that initially have strength only at high momentum.  
The momentum distribution
of the deuteron at large $q$ is our prototype.
The momentum distribution for the initial potentials
in Fig.~\ref{fig:MomDist} show structure at high momenta because of
the short-range repulsion in the potential (particularly for AV18).
When evolved to low momentum, this structure disappears and the
deuteron wave function will select the low-momentum part of the
evolved operator.  But this evolved operator must still reflect the external
high-momentum scale.  We can anticipate simplifications by exploiting
this separation of scales provided by the SRG;
in particular, we expect a factorization of the evolved operator based
on operator-product-expansion (OPE) arguments
applied to nonrelativistic effective theories~\cite{Lepage:1997cs,Braaten:2008uh}.

\subsection{Numerical Verification of Factorization}

\begin{figure}[tbh]
  \subfigure[]{\includegraphics[width=0.45\textwidth]{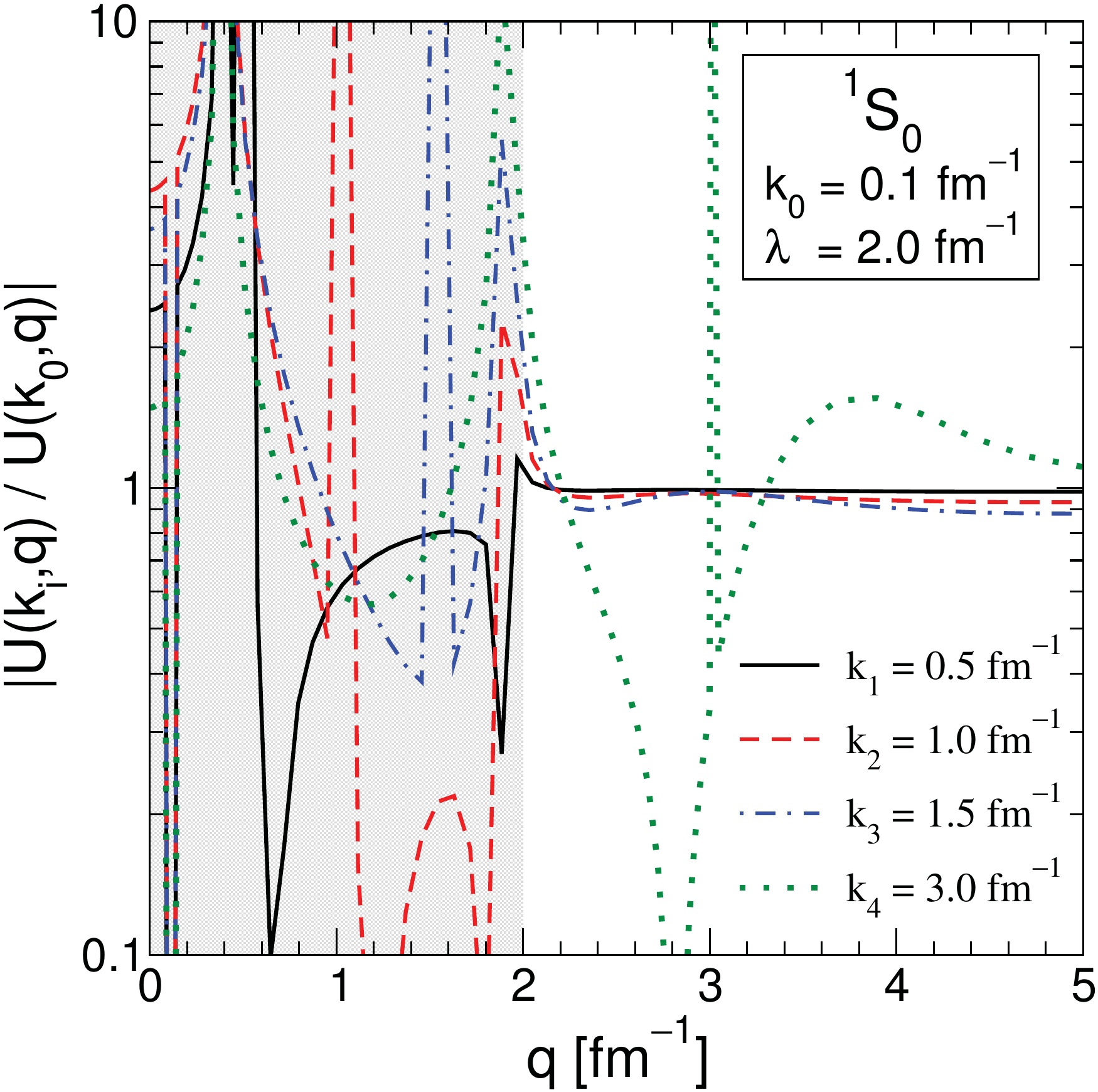}}
  \hspace*{.1in}
  \subfigure[]{\includegraphics[width=0.45\textwidth]{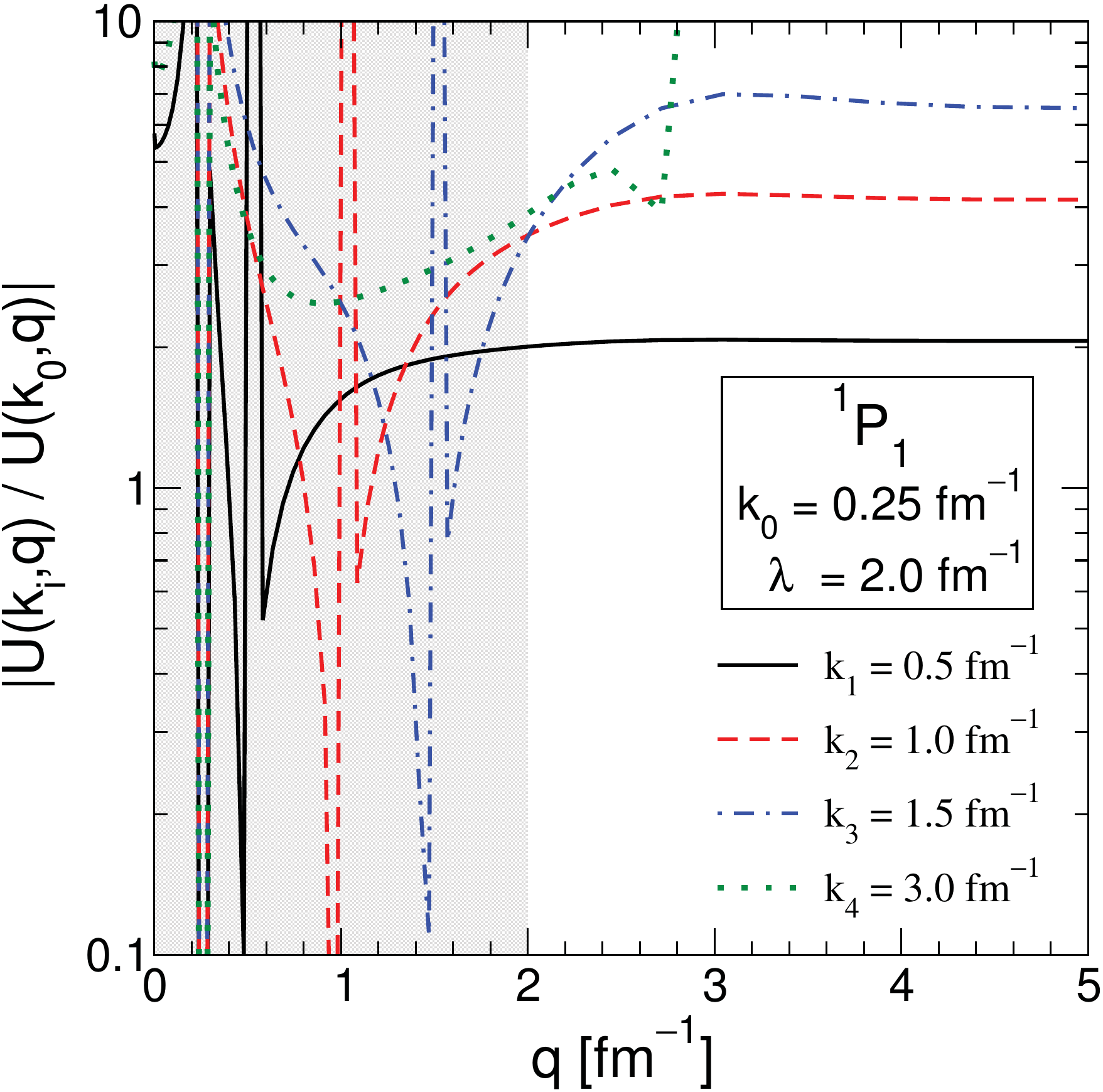}}
  \caption{Numerical tests of factorization of the unitary 
  transformation $U_\lambda(k,q)$ by plotting the ratio in Eqs.~\eqref{eq:Uratio1}
  and \eqref{eq:Uratio2} 
  as a function of $q$ for fixed $k_0$ and several values of $k_i$. 
  Plateaus in $q$ indicate factorization.
  The unitary transformations are generated from the Argonne 
  $v_{18}$ (AV18)~\cite{Wiringa:1994wb} potential evolved to 
  $\lambda = 2\,\mbox{fm}^{-1}$ in the (a) $^1$S$_0$ and 
  (b) $^1$P$_1$ partial waves.  The shaded region marks $q < \lambda$.}
  \label{fig:NumFactorization1}
\end{figure}

\begin{figure}[tbh]
  \subfigure[]{\includegraphics[width=0.45\textwidth]{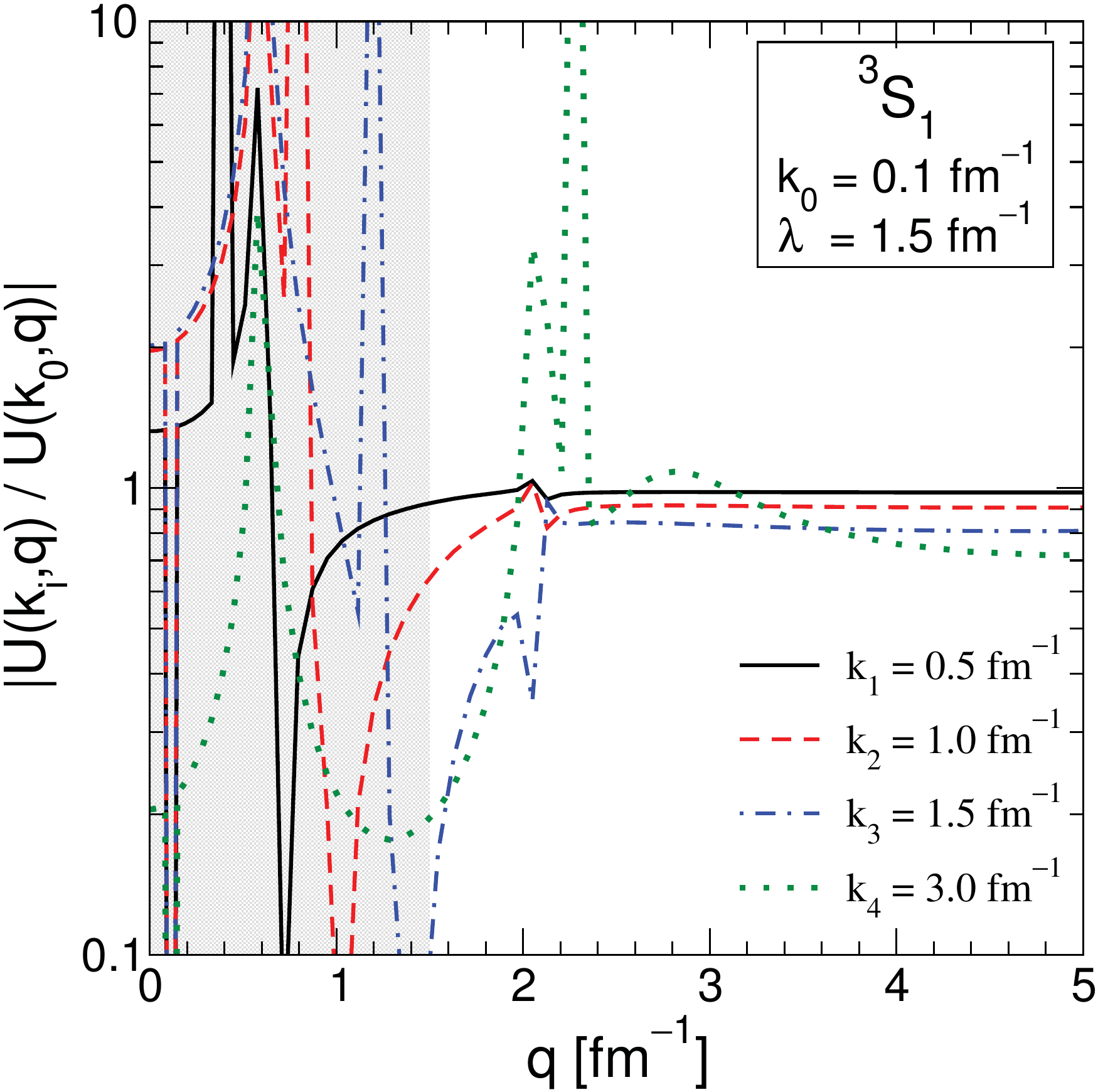}}
  \hspace*{.1in}
  \subfigure[]{\includegraphics[width=0.45\textwidth]{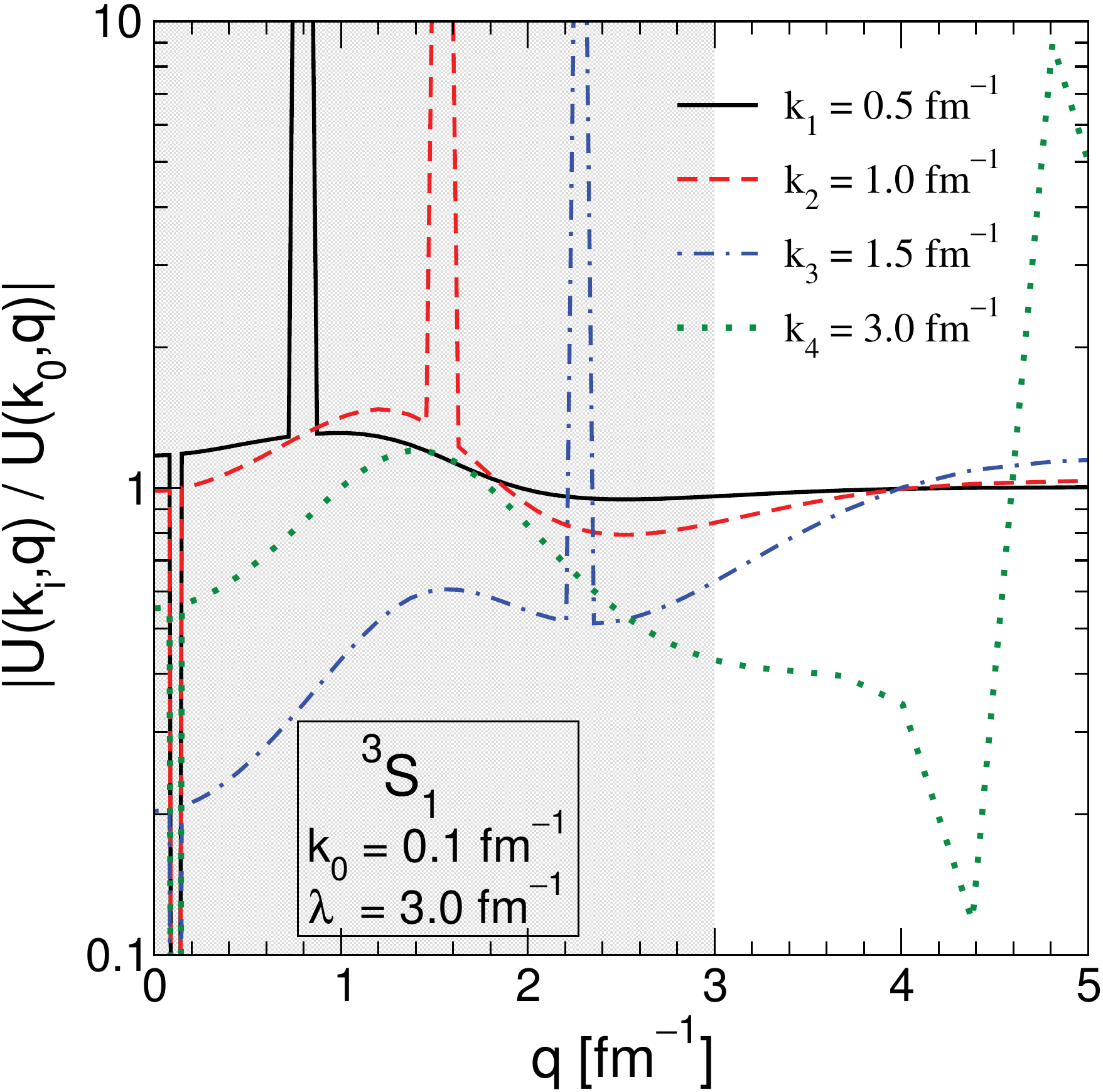}}
  \caption{Same as Fig.~\ref{fig:NumFactorization1} but for the
  $^3$S$_1$ partial wave and $\lambda$ of (a) $1.5\,\mbox{fm}^{-1}$
  and (b) $3\,\mbox{fm}^{-1}$.}
  \label{fig:NumFactorization2}
\end{figure}

Previous calculations of the deuteron momentum distribution suggested
that the unitary evolution operator, \(U_{\lambda}(k,q)\),%
\footnote{Because $\lambda$ is an important momentum scale in
our factorization discussion, we use the notation
$U_\lambda$ with $\lambda = s^{-1/4}$
rather than $U_s$ in this section.} \emph{factorizes} 
into a function of $k$ times a function of $q$, 
\(U_{\lambda}(k,q)\rightarrow K_{\lambda}(k)Q_{\lambda}(q)\),
for  \(k<\lambda\) and \(q\gg\lambda\)~\cite{Bogner:2007jb}. 
To numerically test for factorization in the unitary transformation, 
we use transformations generated via Eq.~\eqref{U:trans} 
and the evolution of $NN$ potentials, and consider the ratio
\beqn
    \frac{U_{\lambda}(k_{i},q)}{U_{\lambda}(k_{0},q)}
    \stackrel{?}{\longrightarrow}
    \frac{K_\lambda(k_{i}){Q_\lambda(q)}}{K_\lambda(k_{0}){Q_\lambda(q)}}
    \;,
    \label{eq:Uratio1}
\eeqn
holding \(k_{i}\) and \(k_{0} \) constant with
\(k_{0}\ll\lambda\).  If there is factorization, the $q$
dependence should cancel; that is, for \(k<\lambda\) and \({q\gg\lambda}\)
we should find
\beqn
  \frac{U_{\lambda}(k_{i},q)}{U_{\lambda}(k_{0},q)}
  \approx
  \frac{K_\lambda(k_{i})}{K_\lambda(k_{0})}
  \;.
  \label{eq:Uratio2}
\eeqn
In Figs.~\ref{fig:NumFactorization1} and \ref{fig:NumFactorization2}
we plot the ratio in Eq.~\eqref{eq:Uratio1} versus $q$ for representative
cases.
The signature of factorization is a plateau in $q$.
The shaded regions are where $q \leq \lambda$.
In all cases, there is no factorization in this region, consistent
with the requirement that $q\gg\lambda$.  In the unshaded
region we see definite plateaus for $q>\lambda$ as long as $k_i < \lambda$,
with diminishing prominence as \(k_{i}\) increases
(they disappear for $k_i>\lambda$).
Thus we have at least a qualitative verification of factorization.
Note that Fig.~\ref{fig:NumFactorization2}(b) shows that
for larger $\lambda$ the clean factorization breaks down (as well
as restricting the applicable domain).

\newcommand{\uvec}{{\bf u}}
\newcommand{\vvec}{{\bf v}}

The singular value decomposition (SVD) can be used as a tool
to quantitatively analyze the extent to which $U_\lambda$
factorizes. The SVD of a matrix $M$ can be
expressed in general as an outer product expansion
\begin{equation}
  M = \sum_{i=1}^{r}d_{i}\uvec_i \vvec_i^{t} 
  \;,
\end{equation}
where $r$ is
the rank of the matrix and the \(d_{i}\) are the singular values (in
order of decreasing value). The idea is that if the first singular
value, \(d_{1}\), is sufficiently large compared to the others, 
the first term dominates and we have a factorized approximation.
We can apply this to $U_\lambda$
in the region where high and low momentum
couple.  Thus,  the vector $\uvec_1$ would correspond to the low
momentum function \(K_{\lambda}(k)\) from Eq.~\eqref{eq:Uratio1}
and $\vvec_1$ to \(Q_{\lambda}(q)\). If valid, one can calculate  the
factorized operator using the unitary transformation obtained
directly from the SVD.   Moreover, the  expansion provided by the
SVD  allows us to make systematic corrections to the factorized
unitary transformation and the operators evolved with it.
\begin{table} 
\begin{tabular}{|c|c|c|c|c|c|c|}\hline
 \mystrut& \multicolumn{3}{c}{\(^1\)S\(_0\)} \vline& 
 \multicolumn{3}{c}{\(^3\)S\(_1\)--\(^3\)D\(_1\)} 
 \vline\\\hline\textbf{Potential} & $ d_1$ & $ d_2$ & $ d_3$  & $ d_1$ & $ d_2$ & $ d_3$\\\hline
AV18 & 0.763 & 0.033 & 0.007  & 0.671 & 0.015 & 0.008\\\hline
N3LO 500 MeV &  1.423 &  0.221 &  0.015  & 1.873 & 0.225 & 0.044\\\hline
N3LO 550/600 MeV & 3.074 & 0.380 & 0.061  & 4.195 & 0.587 & 0.089\\\hline
\end{tabular}
\caption{ Singular values of the unitary transformation \(U(k,q) \)
for  \(q>\Lambda\) and \(k<\Lambda\)  (see discussion in text; units in \(\fmi\)) corresponding to the given
potentials at \(\lambda=2\fmi\) in the \(^1\)S\(_0\) partial wave
and \(^3\)S\(_1\)--\(^3\)D\(_1\) coupled channel.  }
\label{SVD1}
\end{table}

To test if such an expansion can be used, the first few singular
values have been calculated in Table~\ref{SVD1} for the  \(q>\lambda\) and
\(k<\lambda\) region of the SRG unitary transformations
for several different potentials, each evolved
to   \(\lambda=2\fmi\). 
That is, the SVD is applied to the matrix obtained when elements 
of $U_\lambda(k,q)$ with $k>\Lambda$ and $q<\Lambda$ are set to zero; in practice a cutoff \(\Lambda\approx2.5\fmi\) is used to ensure that we are in the region where off-diagonal coupling is strongly suppressed everywhere in the Hamiltonian.
The dominance of $d_1$ in each case is promising.  

To test if
a truncated outer product sum is a good approximation to the contribution
from $k<\Lambda$, $q>\Lambda$,
we consider the errors in some representative 
expectation values in Table~\ref{SVD2}
for several levels of truncation.
The ``zeroth-order'' contribution is from the matrix where
$k>\Lambda$ and $q >\Lambda$ is set to zero (this is denoted by SVD=0 in the table).  The first-order (SVD=1 in the table)  contribution uses the same matrix plus the approximation of    
$U_\lambda(k,q)$ for $k<\Lambda$ and $q>\Lambda$ given by the first outer product in the SVD\ expansion.  The second order approximation uses two outer products, etc.  The occupation and charge form factor operators shown here are chosen to illustrate the effects of the factorized approximation at various momenta.  Additionally, the  initial occupation operator has no off-diagonal strength, whereas the initial charge form-factor operators have relatively substantial off-diagonal contributions at large values of \(q \); this is significant for the applicability of factorization to an operator, as we see below. So, what we find at low momentum for the occupation operator  is that
it is essentially exact, up to errors resulting from decoupling and truncation of the wave function, and it is the same with or without the SVD approximation.  This error increases, as expected, for larger \(\lambda\). Because \(G_{C} \) is more diffuse initially, we see a small improvement even at small momenta when using the SVD approximations.  

For the occupation operator at high momenta (well above the cutoff), the error is 100\% without an approximation to \(U_\lambda(k,q)\)  because there is a hard cutoff and the initial operator is localized in the upper region of momentum space. However, with just one term in the SVD\ expansion we recover that expectation value for \(\lambda=1.5\fmi\) to better than 1\%, and the situation improves with additional terms in the expansion.  At \(\lambda=3.0\fmi\) decoupling is evidently not sufficient for this approximation to work well. At very large values of momenta (e.g., \(q\approx6.9 \fmi\)) the charge form factor shows improvement with the SVD\ approximation, but because this operator has significant off-diagonal strength, the improvement is not as pronounced.   At  a value of  \(q\approx3.0 \fmi\) it is evident that the SVD still improves the relative error.  However, recall that the strength in the form factors is larger around \(\approx\frac{1}{2}q\). \   \ 

\begin{table}[tbp]
\begin{tabular}{|c|c|c|c|c|c|c|c|c|}\hline
 & \multicolumn{4}{c}{\mystrut SRG $\lambda=1.5\fmi$} \vline& \multicolumn{4}{c}{SRG $\lambda=3.0\fmi$} \vline\\\hline \textbf{Operator} & SVD & $q=0.34$ & $q=3.01$ & $q=6.90$& SVD  & $q=0.34$ &$q=3.01$ &$q=6.90$\\\hline 
\(\langle a^{\dagger}_{q}\,a_q\rangle\)    & \mystrut0 & 7.61\(\times10^{-7}\) & 1.00&  & 0 & 1.06\(\times10^{-3}\) & 1.00&\\\hline
~with N$^3$LO~ &  \mystrut1 &  7.61\(\times10^{-7}\)  &  4.28\(\times10^{-3}\)&  & 1 & 1.06\(\times10^{-3}\) & 6.36\(\times10^{-1}\)&\\\hline
\mystrut & 2 & 7.61\(\times10^{-7}\) & 4.79\(\times10^{-4}\)&  & 2 & 1.06\(\times10^{-3}\) & 6.35\(\times10^{-1}\)&\\\hline
\hline
\(G_{\rm C}(q) \)     & \mystrut0 & 6.90\(\times10^{-4}\) & 5.01\(\times10^{-3}\) &8.93\(\times10^{-1}\) & 0 & 4.10\(\times10^{-4}\) & 3.36\(\times10^{-3}\)&8.92\(\times10^{-1}\) \\\hline
~with NNLO~ &  \mystrut1 &  1.28\(\times10^{-7}\) &  8.90\(\times10^{-5}\) & 4.06\(\times10^{-2}\)& 1 & 1.63\(\times10^{-4}\) & 2.66\(\times10^{-4}\)&4.00\(\times10^{-1}\)\\\hline
 & \mystrut2 & 1.04\(\times10^{-6}\) & 2.10\(\times10^{-5}\)&4.18\(\times10^{-2}\)  & 2 & 1.63\(\times10^{-4}\) & 3.04\(\times10^{-4}\)&4.09\(\times10^{-1}\)\\\hline
\end{tabular}
\caption{ Relative error of evolved operator matrix elements calculated using the SVD to factorize $U_\lambda(k,q)$ in the region where $k<\Lambda$ and  $q>\Lambda$ (see discussion in text; units in \(\fmi\)).  }\label{SVD2}
\end{table}

\subsection{Connection to the Operator Product Expansion}

\newcommand{\Hinf}{H_{\infty}}
\newcommand{\Vinf}{V_{\infty}}
\newcommand{\psilamalpha}{\psi^\lambda_\alpha}
\newcommand{\psiinfalpha}{\psi_\alpha^{\infty}}

The OPE was developed for the evaluation of singular products of local field operators at small separation. In our case, where such operators are treated as matrices and we typically work in momentum representation, the focus becomes 
low-momentum matrix elements of a product in which high-momentum states dominate the intermediate sum.
This leads us directly to consider low- to high-momentum matrix elements of SRG-evolved operators, and a generic analysis is then based on the study of $U_\lambda(k,q)$ for $k<\lambda$ and $q\gg\lambda$.

The utility of the OPE rests on factorization; short-distance details decouple from long-distance dynamics. Factorization enables one, for example, to separate the momentum and distance scales in hard-scattering processes in terms of perturbative QCD and parton distribution functions. In our case, factorization is the direct result of decoupling. It provides tools that let us parametrize the high-momentum components of operators that would normally require degrees of freedom we do not retain. We can, for example, build effective few-body operators containing state-independent functions of high momenta that can be measured directly in few-body experiments. These operators can then be employed to make predictions for $A$-body systems.

Consider a generic operator, $\Oop_\lambda = U_\lambda \Oop U^\dagger_\lambda$, and employ the spectral representation for $U_\lambda$:
\beqn
U_\lambda (k,q) = \sum_\alpha \langle k|\psi_\alpha(\lambda)\rangle \langle \psi_\alpha(\infty)|q\rangle .
\eeqn
The OPE deals with cases in which the unevolved operator is dominated by high momenta (e.g., $a^\dagger_q a_q$ with large \(q\) is the simplest paradigm) and we focus on $k<\lambda$ and $q\gg\lambda$. For $k<\lambda$ we exploit the fact that low-momentum components of high-energy eigenstates of $H_\lambda$ are exponentially suppressed because of decoupling. As a result the sum is dominated by low-energy states,
\beqn
U_\lambda (k,q) \approx \sum_{E_\alpha\ll\lambda^2} \langle k|\psi_\alpha(\lambda)\rangle \langle \psi_\alpha(\infty)|q\rangle .
\eeqn

Once the sum is restricted we can turn our focus to approximating the high-momentum components of the unevolved low-energy states. This analysis is closely related to Lepage's discussion of the OPE analysis of wave functions, which leads him to write for S-waves in position representation~\cite{Lepage:1997cs}:
\beqn
  \Psi_{\rm true}(r) = \overline{\gamma}(r) \int\! d^{3}r\,
      \Psi_{\rm eff}\delta^{3}_{a}(\boldsymbol r)
   + \overline{\eta}(r)a^{2}\int\! d^{3}r\,
   \Psi_{\rm eff}\nabla^{2}\delta^{3}_{a}(\boldsymbol r)
   + \mathcal{O}(a^{4})
   \;,\label{eqn:Lepage}
\eeqn
where the coefficient functions
\(\overline{\gamma}(r)\) and \(\overline{\eta}(r)\) are
state-independent parametrizations of the short-distance physics, and \(a\) is approximately the distance of the ultraviolet cutoff. In this section, we outline how SRG factorization
can be understood
more generally (and analytically) in the context of the OPE
for nonrelativistic Schr\"odinger problems by deriving an analogous equation in momentum space for the SRG-evolved wave function. 

To do so, we first define the projection operators 
\beqn
  \projP_{\Lambda} = \int^{\Lambda}_{0}\! d\wt p \ \vert p 
  \rangle\langle p \vert 
\eeqn
and 
\beqn
  \projQ_{\Lambda} = \int^{\infty}_{\Lambda}\! d\wt q \
    \vert q \rangle\langle q \vert 
    \;,
\eeqn
where \(\Lambda\) divides momentum space and \(d\wt p\equiv \frac{2}{\pi}p^{2}\,dp\) in the partial-wave momentum basis.
This \(\Lambda\) is to be
distinguished from \(\lambda\), which is the SRG evolution parameter
and  an approximate measure of decoupling in the evolved potential. 
We use \(\psilamalpha\) to denote the
eigenstates of the Hamiltonian ordered according to
increasing energy \(E_{\alpha}\) and evolved to \(\lambda\) via the SRG. 
\(H_{\lambda} \) and \(V_{\lambda}\) represent the corresponding SRG
evolved Hamiltonian and potential. 
The initial, unevolved  operators correspond to \(\lambda=\infty\).

%omitting the subscript $\lambda$.

From the unevolved Schr\"odinger equation
\beqn
  \Hinf\left|\psiinfalpha \right>
    = E_{\alpha}\left|\psiinfalpha \right\rangle
    \;,
\eeqn
 we can write
\beqn
  \begin{pmatrix}
    \projP_{\Lambda}\Hinf\projP_{\Lambda} & 
    \projP_{\Lambda}\Hinf\projQ_{\Lambda} \\
    \projQ_{\Lambda}\Hinf\projP_{\Lambda} & 
    \projQ_{\Lambda}\Hinf\projQ_{\Lambda} \\
  \end{pmatrix}
  \begin{pmatrix}
    \projP_{\Lambda}\psiinfalpha\ \\
    \projQ_{\Lambda}\psiinfalpha \\
  \end{pmatrix}
    = E\\ _{\alpha}\begin{pmatrix}\projP_{\Lambda}\psiinfalpha \\
        \projQ_{\Lambda}\psiinfalpha \\
  \end{pmatrix}
  \;,
\eeqn
and thus for the ``$\projQ$" space we have
\bea
  \projQ_{\Lambda}\left\vert \psiinfalpha\right>
  &=& (E_{\alpha}-\projQ_{\Lambda}\Hinf\projQ_{\Lambda})^{-1}
  \projQ_{\Lambda}\Hinf\projP_{\Lambda}
  \projP_{\Lambda}\left\vert \psiinfalpha \right\rangle
  \nonumber \\
    &=& (E_{\alpha}
     -\projQ_{\Lambda}\Hinf\projQ_{\Lambda})^{-1}\projQ_{\Lambda}\Vinf
     \projP_{\Lambda}\left\vert \psiinfalpha \right\rangle  
  \;,
\eea
where we have used
\((\projP_{\Lambda})^{2}=\projP_{\Lambda}\), \(\Hinf=T+\Vinf\),
and \(\projQ_{\Lambda}T\projP_{\Lambda}=0\).
For low-energy states \( \psiinfalpha\) such that
 \(|E_{\alpha}|\ll  {\rm Min}[|E_{\projQ H\projQ }|]\) (where \(E_{\projQ H\projQ }\) are the eigenvalues of \(\projQ H\projQ\)), we can neglect the \(E_{\alpha}\)
 dependence.  Also, assuming that the  potential
 \(\Vinf(q',p)\) is slowly varying with respect to \(p\) compared to 
 \( \psiinfalpha(p)\) in the region \(p<\Lambda\) and
 \(q'\gg\Lambda\), 
 we can use the expansion for S-waves %
\begin{eqnarray}\nonumber
   \int^{\Lambda}_{0} \! d\wt p \,
     \Vinf(q',p)\psiinfalpha (p)
      &\approx& 
      \Vinf(q',p')\vert_{p'=0}\times 
        \int^{\Lambda}_{0}\! d\wt p \,\psiinfalpha (p)
      \\  && 
      \null + \left.\frac{d^{2}}{dp^{'2}}\Vinf(q',p')
      \right\vert_{p'=0}\times \int^{\Lambda}_{0}\! d\wt p\,p^{2}\,\psiinfalpha (p)\,+\cdots
\label{eq:Taylor}
\end{eqnarray}
 to first order, combined with the fact that the low-energy states
will have momentum components peaked at small $p$, to  write 
\beqn
  \left\langle q|\psiinfalpha 
    \right\rangle\approx-\int^{\infty}_{\Lambda}\! d\wt{q}'
    \int^{\Lambda}_{0}\! d\wt p \left\langle q\right\vert
    \frac{1}{\projQ_{\Lambda}\Hinf\projQ_{\Lambda}}\left| q' \right> 
    \Vinf(q',0)\,\psiinfalpha(p) \;.
\eeqn
Tests indicate that these assumptions are valid for realistic $NN$
potentials.   

Further, we see empirically via Fig.~\ref{fig:MomDist} 
that \(\projP_{\Lambda}\left\vert
\psiinfalpha \right\rangle\approx Z(\lambda)\left\vert
\psilamalpha \right\rangle\) when \(\lambda\gtrsim\Lambda\) 
(this is consistent with our understanding that the SRG with
\(G_s=\trel\) renormalizes/suppresses only the short-distance
components of the wave function for values of 
\(\lambda\) considered here).
Thus, setting \(\Lambda=\lambda\) and defining
\beqn
\gamma ^\lambda(q)\equiv-\int^{\infty}_{\lambda}d\wt{q}'\,
 \left\langle q\right\vert
 \frac{1}{\projQ_{\lambda}\Hinf\projQ_{\lambda}}\left| q' \right> \Vinf(q',0)
 \;,
 \label{eq:gammalambda}
\eeqn
we have 
\beqn
   \psiinfalpha(q) \approx  \gamma^{\lambda}(q)
           \int^{\lambda}_{0}\! d\wt p\, Z(\lambda)\psilamalpha (p)
   \;.
   \label{eq:psialphaq}
\eeqn 
So we see that the high-momentum components of low-energy
eigenstates can be factorized into a state-independent function
\(\gamma ^\lambda(q)\), which summarizes the short distance behavior
of the wave function, and a coefficient (given by an integral over
the renormalized wave function) that gives the contribution due to
the long-distance structure of the state.  Moreover, if we include
higher-order corrections resulting from the expansion of
\(\int^{\lambda}_{0}\! d\wt p\,\Vinf(q',p)\psiinfalpha (p)\)
about \(p=0\), we recover the analog to Lepage's OPE, Eq.~\eqref{eqn:Lepage}, in momentum space for the short-distance structure of a  wave function.  It is given by
\beqn
 \psiinfalpha(q) \approx \,\gamma^{\lambda}(q)\int^{\lambda}_{0}\! d\wt p\, 
  Z(\lambda)\psilamalpha (
  p)+\eta^{\lambda}(q)\int^{\lambda}_{0}\!d\wt p\,p^{2}\,
  Z(\lambda)\,\psilamalpha (p)+\cdots
\eeqn
where \(\gamma^{\lambda}(q)\) is given previously and
\beqn
   \eta^{\lambda}(q) \equiv 
     -\int^{\infty}_{\lambda}\! d\wt{q}{}'\,
     \left\langle q \right\vert
     \frac{1}{\projQ_{\lambda}\Hinf\projQ_{\lambda}}\left\vert 
     q' \right> \left.\frac{\partial^{2}}{\partial p^{2}}\Vinf(q',p)\right|_{p=0}
     \;.
\eeqn 

Now, from the definition of  the SRG unitary evolution operator
in Eq.~\eqref{U:trans}, in the region
\(k<\lambda\) and \(q\gg\lambda\) we can use the leading-order
term of our OPE to write
\begin{eqnarray}
  U_{\lambda}(k,q) &=&
   \sum^{\infty}_{\alpha}\left\langle k| \psilamalpha\right\rangle
   \left\langle \psiinfalpha | q \right\rangle
   \nonumber \\ &\approx&
   \left[\sum^{|E_{\alpha}| \ll  |E_{\projQ H\projQ }|}_{\alpha}
   \left<k| \psi^{\lambda}_{\alpha}\right\rangle 
   \int^{\lambda}_{0}\! d\wt p\, Z(\lambda)\psilamalpha{}^\dag(p)\right] 
   \gamma^{\lambda}(q)
   \nonumber \\  &\equiv& K_{\lambda}(k)Q_{\lambda}(q)
   \;,\label{factorizeU:derived1}
\end{eqnarray}
where the sum is only over states in the ``\(\projP\)" space thanks to decoupling. Thus,  we can understand 
the factorization of \(U_{\lambda}\) as a general consequence of our
ability to factorize the high-momentum components of low-energy
nuclear wave functions via an OPE plus  decoupling in the 
SRG-evolved Hamiltonian.  Moreover, because $\psilamalpha(k)$ to good
approximation has
no support when $k < \lambda$ for $\alpha$ in the ``$\projQ$''
space, we can extend the $\alpha$ sum in Eq.~\eqref{factorizeU:derived1}
to the full space
and apply closure to find
\beqn
  U_{\lambda}(k,q) \approx
   \left[
   Z(\lambda)
   \int^{\lambda}_{0}\! d\wt p\sum^{\infty}_{\alpha}
   \left<k| \psi^{\lambda}_{\alpha}\right\rangle \left<\psi^{\lambda}_{\alpha}| p\right\rangle \right] 
   \gamma^{\lambda}(q)
   \approx Z(\lambda)\gamma^{\lambda}(q)
   \;.
   \label{eq:leadingfactorized}
\eeqn
Thus, to a first approximation, \(K_{\lambda}(k)\) is a constant factor. 

This approximate constancy implies that the ratios for the $L=0$
channels in Figs.~\ref{fig:NumFactorization1}(a)
and \ref{fig:NumFactorization2}(a) and (b) should tend to one
in the factorization region, which is realized at the 10--20\%
level for sufficiently low $\lambda$.
For $L>0$, the generalization of Eq.~\eqref{eq:leadingfactorized}
follows from modifying the Taylor series in Eq.~\eqref{eq:Taylor}
to account for $V_\infty(q',p) \propto p^L$ for small $p$.
Then Eqs.~\eqref{eq:gammalambda} and \eqref{eq:psialphaq}
are changed to
\beqn
\gamma ^\lambda(q)\equiv-\int^{\infty}_{\lambda}d\wt{q}'\,
 \left\langle q\right\vert
 \frac{1}{\projQ_{\lambda}\Hinf\projQ_{\lambda}}\left| q' \right> 
 \left.\frac{d^L}{dp^L}\Vinf(q',p)\right\vert_{p=0} 
 \;,
 \label{eq:gammalambdaL}
\eeqn
and
\beqn
   \psiinfalpha(q) \approx  \gamma^{\lambda}(q)
           \int^{\lambda}_{0}\! d\wt p\, Z(\lambda) p^L \psilamalpha (p)
   \;.
   \label{eq:psialphaqL}
\eeqn 
With these changes, Eq.~\eqref{eq:leadingfactorized}
for the factorization at leading approximation becomes
\beqn
  U_{\lambda}(k,q) \approx
   \left[
   Z(\lambda)
   \int^{\lambda}_{0}\! p^L d\wt p\sum^{\infty}_{\alpha}
   \left<k| \psi^{\lambda}_{\alpha}\right\rangle \left<\psi^{\lambda}_{\alpha}| p\right\rangle \right] 
   \gamma^{\lambda}(q)
   \approx k^L Z(\lambda)\gamma^{\lambda}(q)
   \;.
   \label{eq:leadingfactorizedL}  
\eeqn
This approximation implies that the ratios in the factorization
region should tend for $L>0$ to $(k_i/k_0)^L$, which is
seen at the same 10--20\% level in Fig.~\ref{fig:NumFactorization1}(b).

To gain insight into the  implications of this
factorization, we consider 
the   expectation value of $\adaggera$ in a low-energy state,
the deuteron.
Because we know that strength in the evolved  number
operator expectation value decouples from high-momentum
contributions in the deuteron, we can write
\begin{eqnarray}\nonumber
  \left\langle \psi^{\lambda}_d\right\vert 
    \left(a_{ q}^{\dagger}a_{q}\right)_{\lambda}
   \left\vert \psi^{\lambda}_d\right\rangle 
   &=& \left\langle   \psi^{\lambda}_d\right\vert  
   U_{\lambda}\left(a_{ q}^{\dagger}a_{q}\right)
   U^{\dagger}_{\lambda}\left\vert \psi^{\lambda}_d\right\rangle 
   \\  \nonumber
   &\approx&     \int^{\lambda}_{0}\!d\tilde {k'} \int^{\infty}_{0}\! d\tilde {q}'
      \int^{\infty}_{0}\! d\tilde q''\int^{\lambda}_{0}\! d\tilde {k}
      \, \psi^{\lambda}_d{}^\dagger(k')U_{\lambda}(k',q')
      \delta (q'-q)
      \delta (q''-q')U_{\lambda}(q'',k)\psi^{\lambda}_d(k)
\\
   &=& \int^{\lambda}_{0}\! d\wt {k'} \int^{\lambda}_{0}\! d\wt{k}\;
    \psi^{\lambda}_d{}^\dagger(k')U_{\lambda}(k',q)U_{\lambda}(q,k)
    \psi^{\lambda}_d(k).
\label{NumOp:ExpectVal}
\end{eqnarray}
For a low-momentum operator, one with \(q<\lambda\), the expectation
value thus depends only on the low-momentum details of the wave
function (original and evolved).  For  \(q\gg\lambda\), however, we
can make use of factorization and set    \(U(k,q)\rightarrow
K_{\lambda}(k)Q_{\lambda}(q)\) to write 
\beqn
   \int^{\lambda}_{0}\! d\wt{k'}\int^{\lambda}_{0}\!d \wt k\,
   \psi^{\lambda}_d{}^\dagger(k')K_\lambda(k')
   [Q_\lambda(q)Q_\lambda(q)]K_\lambda(k)\psi^{\lambda}_d(k)
\eeqn
from Eq.~(\ref{NumOp:ExpectVal}).    Here we see that the
expectation value of a high-momentum number operator is independent
of the long-distance structure of the wave function.  This is
consistent with earlier calculations of the deuteron momentum
distribution \cite{Bogner:2007jb}.  Again, as with decoupling in the
potential, we appear to have a means by which long- and 
short-distance details can be separated for an operator evolved via the  
SRG.

The generalization of this result is straightforward.
Consider the   
expectation value of an arbitrary operator,  \(O(q',q)\),
in a low-energy state, \(\psi^{\lambda}_{\rm low}\). 
Because decoupling is valid for operator
expectation values in a momentum basis (as we have seen via the
expectation value integrand plots  in 
Secs.~\ref{sec:DeuteronExpectVal} and \ref{sec:otherops}), we can write
\beqn
 \left\langle \psi^{\lambda}_{\rm low}\right\vert 
   U_{\lambda}\Oop\,
  U^{\dagger}_{\lambda}\left\vert \psi^{\lambda}_{\rm low}\right\rangle 
  \approx  
  \int^{\lambda}_{0}\! d\wt{k}'
  \int^{\infty}_{0}\! d\wt{q}'\int^{\infty}_{0}\! d\wt q
  \int^{\lambda}_{0}\! d\wt k\,
  [\psi^{\lambda}_{\rm low}(k')]^\dagger
  U_{\lambda}(k',q')
  O(q',q) U_{\lambda}(q,k) \psi^{\lambda}_{\rm low}(k)
  \;.
\eeqn
We separate the integrals over the operator in the expectation
value and apply factorization to set  
\(U(k,q)\rightarrow K_{\lambda}(k)Q_{\lambda}(q)\)
in the region where \(k<\lambda\) and \(q\gg\lambda\).   
If the \textit{unevolved} operator has coupling between high
and low momentum above the the factorization cut, then
there is no great simplification.
However, if the
\textit{unevolved}  operator \textit{does not} have coupling of high
and low momentum above the  factorization cut, factorization will
allow us to separate out the high- and low-momentum structure of an
operator into two contributions:
\beqn
  \int^{\lambda}_{0}\! d\wt{k}' \int^{\lambda}_{0}\! d\wt{q}'
  \int^{\lambda}_{0}\! d\wt q\int^{\lambda}_{0}\!d\wt k\,
  [\psi^{\lambda}_{\rm low}(k')]^\dagger
  U_{\lambda}(k',q')O(q',q)U_{\lambda}(q,k)
  \psi^{\lambda}_{\rm low}(k)
\eeqn
and
\beqn
  \int^{\lambda}_{0}\! d\wt{k}' \int^{\infty}_{\lambda}\! d\wt{q}'
  \int^{\infty}_{\lambda}\! d\wt q \int^{\lambda}_{0}\! d\wt k\,
  [\psi^{\lambda}_{\rm low}(k')]^\dagger K_\lambda(k')[Q_\lambda(q')
  O(q',q)Q_\lambda(q)]K_\lambda(k)\psi^{\lambda}_{\rm low}(k)
  \;.
\eeqn
This is analogous to what was found for the number operator.  

Thus, we see that the  breakdown of contributions to the expectation
value of a general operator is consistent with our
interpretation of the SRG\ flow equations as a means by which one
can achieve a separation of scales in the evaluation of nuclear
few- and many-body problems.  We see explicitly here
that the effects of a low-momentum probe of the ground state wave
function depends  (almost entirely) on the low-momentum details of
the renormalized wave function.  Likewise, the effect of a 
high-momentum probe is largely independent of the low-momentum
structure.   It is only for operators which probe the coupling of
high and low momentum (long and short distance) details of the wave
function that we must consider the full momentum space evolution of
the operator. For the operators that have been
considered in this paper, the latter is only true of the electromagnetic form
factors at relatively high momenta (beyond the typical regime
of interest for nuclear structure); for any operators which weakly
couple high and low momentum, these terms can be neglected.

To summarize, we can write the expectation value of an operator that
has weak coupling between high and low momentum as 
\begin{eqnarray}\nonumber
   \left\langle \psi^{\lambda}_{\rm low}\right\vert 
   U_{\lambda}\Oop\,U^{\dagger}_{\lambda}\left\vert 
   \psi^{\lambda}_{\rm low}\right\rangle 
   &\approx& \int^{\lambda}_{0}\!d\wt{k'}\int^{\lambda}_{0}\!d\wt k \,  
   [\psi^{\lambda}_{\rm low}(k')]^\dagger
    \biggl[\int^{\lambda}_{0}\!d\wt{q'}\int^{\lambda}_{0}\!d\wt q \;
  \nonumber \\
  & & \quad\null \times   
  U_{\lambda}(k',q')O(q',q)U_{\lambda}(q,k)
    +I_{QOQ}K_{\lambda}(k')K_{\lambda}(k)
   \biggr] \psi^{\lambda}_{\rm low}(k)
   \;,
\label{eq:Factorizedexpectation}
\end{eqnarray}
where
\beqn
  I_{QOQ} = \int^{\infty}_{\lambda}\!d\wt{q'}\int^{\infty}_{\lambda}\!d\wt q 
             ~Q_{\lambda}(q')O(q',q)Q_{\lambda}(q) \;.
   \label{eq:UniversalHighMom}
\eeqn
 By using factorization, we have seen that the expectation value
breaks into a sum of two components: one which describes 
low-momentum structure, and a high-momentum component that factorizes
into a piece depending on the low-momentum structure, and another piece
that is a universal function of high momentum $q$.

\subsection{Interpreting high-momentum scaling behavior of the momentum 
  distribution}  \label{subsec:interpret}

As discussed in Sec.~\ref{sec:intro}, results from
large-momentum-transfer experiments
such as $(e,e'p)$ have been related \cite{Frankfurt:2008zv}
to how the tails
of momentum distributions calculated 
for nuclei and nuclear matter using phenomenological potentials exhibit
a scaling behavior at high momentum~\cite{Pieper:1992gr}. 
In particular,
the momentum dependence of the distributions for nuclei ranging from
the deuteron to oxygen, as well as nuclear matter, at high momentum
is similar except for an overall nucleus-dependent scaling factor.  
One explanation for this is based on
the dominance of two-body forces in the
interaction and  short-range correlations in the wave 
functions~\cite{Frankfurt:2008zv}.
How can we explain this feature in an SRG-evolved calculation,
for which high-momentum components and short-range correlations
are suppressed?

Factorization provides a compelling alternative explanation.
Because the $\adaggera$ operator for large $q$ has no coupling to low
momentum, the entire $q$ dependence comes through the
function $I_{QOQ}$ in Eqs.~(\ref{eq:Factorizedexpectation}) and (\ref{eq:UniversalHighMom}), 
which is independent of the low-momentum part.
If induced many-body contributions to the operator are relatively
small and we neglect for the moment effects from embedding two-body operators into an $A > 2$ space, 
we conclude that for $A \geq 2$  the momentum distribution
should be approximately the same for every $A$, 
with a scaling factor given by an $A$-dependent low-momentum
integral over the low-energy wave functions.

\begin{figure}[tbh]
  \includegraphics[width=.75\textwidth,clip]{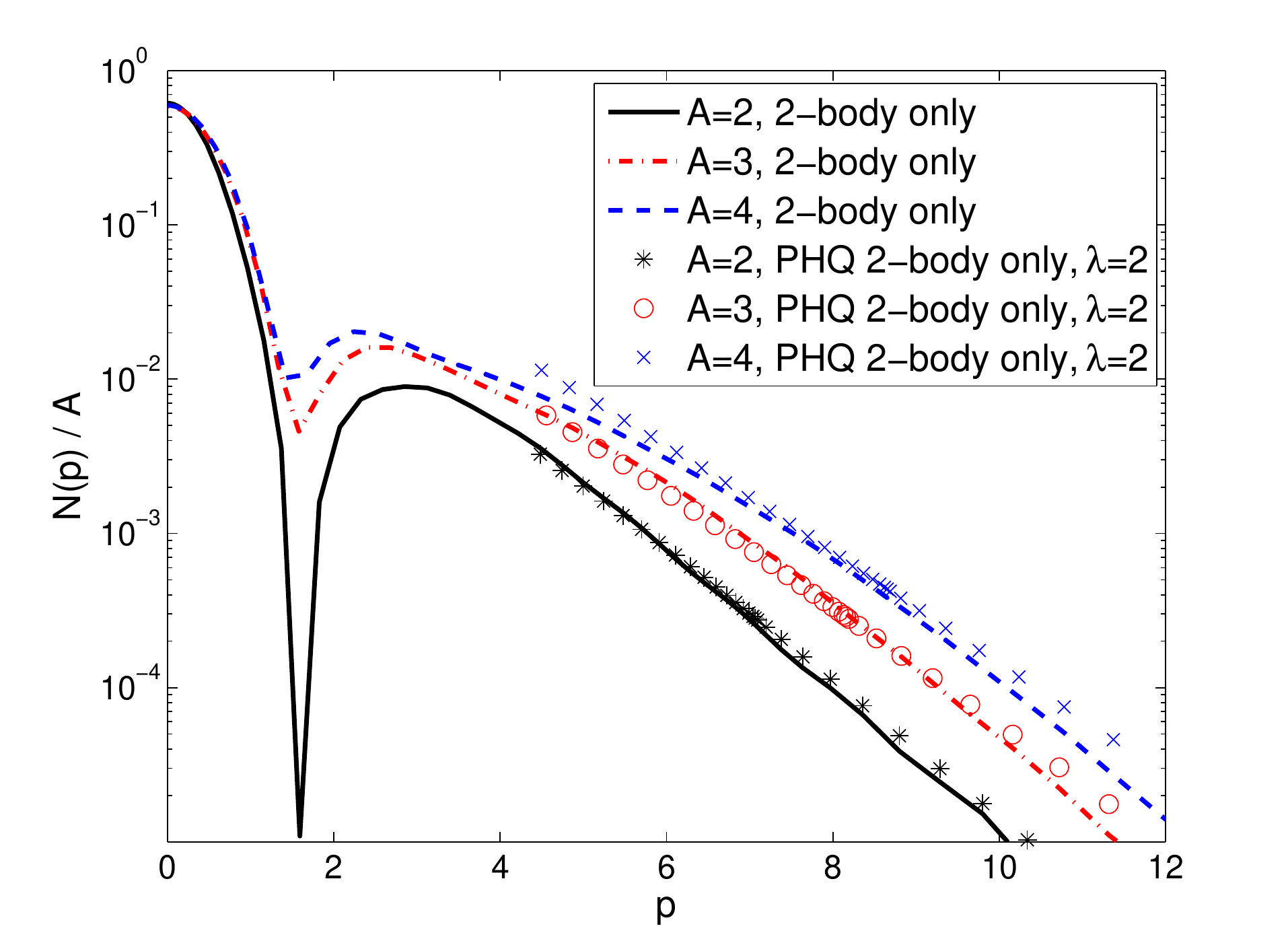}
  \caption{The scaling of momentum distributions at high momenta in a
  1D model is tested by using the leading-order factorized approximation
  to the momentum occupation number operator to predict high-momentum
  scaling in $A=2,3,4$ (symbols).  
  The full momentum distributions for $A=2$, 3, and 4
  are shown with solid, dot-dashed, and dashed lines, respectively.}  
\label{fig:FactorizationScaling}
\end{figure}

We have tested this proposal in a 1D model with an interaction that
mimics features of the nuclear $NN$ potential.  The
model and procedures used for the 1D calculations are described in
Ref.~\cite{Jurgenson:2008jp}.  The full momentum distributions  for two,
three, and four particle systems in this model are shown with solid,
dot-dashed,
and dashed lines, respectively, in Fig.~\ref{fig:FactorizationScaling}.  
The behavior at
high momentum is analogous to the nuclear calculations~\cite{Pieper:1992gr}: 
The momentum dependence is similar for each system so that each curve
differs only by a scaling factor.  We then evolve the model
interaction via the SRG to \(\lambda=2\) and extract the unitary
transformation.  
Only the operator from $A=2$ is used;
that is, any induced three- or four-body component is neglected.
Details of the extraction and embedding of operators for $A>2$,
including boost corrections, are described in the sequel
\cite{Anderson:2010aa}.  
By using the first term in the SVD expansion to factorize \(U(k,q)\) in the region where \(k<\lambda\) and \(q\gg\lambda\), we are able to reproduce to a large extent the momentum
distributions at high momenta (shown with symbols in 
Fig.~\ref{fig:FactorizationScaling}), and confirm our expectations regarding the scaling of the curves. These results are very promising and merit further investigation.

%%%%%%%%%%%%%%%%%%%%%%%%%%%%%%%%%%%%%%%%%%%%%%%%%%%%%%%%%%%%%%%%%%%%%%%%%%
%%%%%%%%%%%%%%%%%%%%%%%%%%%%%%%%%%%%%%%%%%%%%%%%%%%%%%%%%%%%%%%%%%%%%%%%%%
\section{Summary}  \label{sec:summary}

In this article, we have examined the evolution of operators via the
SRG with restricted application to the
deuteron.
We considered only the most commonly used generator $G_s = [T_{\rm rel},H_s]$
with normal ordering in the vacuum.
At this two-particle level it is easy to ensure that the transformations 
are unitary to high accuracy, so the invariance of matrix elements 
is assured.  Thus our focus is instead on the nature of the evolved
operators:  Does a form of decoupling apply?  Do operators become increasingly
complicated as the wave functions become increasingly less correlated?
How large are induced two-body contributions to various one-body operators?

By considering the operator matrix elements in momentum
representation both with and without deuteron wave functions included,
we are able to follow the flow of strength.
Because the transformations are unitary,
the integrated value does not change with $\lambda$, but the nature
of the operator does. 
There is little evolution in long
distance operators, whereas high-momentum operators must evolve
significantly to compensate for suppression of  high momenta in 
low-energy wave functions.  In the end, one can see that the movement of the
strength in the   operator expectation values  is given by the
evolution of the eigenstates of the Hamiltonian itself.  Moreover,
we find that decoupling succeeds for operator expectation values in
general, not just for the binding energies.    

The momentum distribution is particularly interesting because the
evolution of high-momentum operators leads to their strength flowing
completely to low momentum.  Thus while the deuteron wave function has rapidly decreasing
support at the high momentum, its matrix elements are preserved 
without pathologies in the transformed operators.  Indeed, operator matrix
elements are less sensitive to details, as evidenced by the improved
effectiveness of estimates using variational wave functions.
Decoupling for operators follows as the contributions from higher energy/momentum
basis states become unimportant, allowing truncation.
This was explicitly illustrated in Figs.~\ref{fig:MDdecouplingAV18}
and \ref{fig:MDdecouplingN3LO}.
The generality of these conclusions is evident by considering the eigenvector
expansion of the SRG unitary transformations, which dictates the flow of
strength.

For low-momentum operators, which also includes the low-momentum part of 
one-body electromagnetic form factors, there is relatively little 
running and therefore only small induced
two-body parts (which for $A=2$ is simply the difference between the initial 
and the evolved result). In general, if the initial operator matrix
elements pick up their strength predominantly at long distance,
the operators will evolve only slightly until $\lambda$ is small.
For electroweak operators, the real interest is in few- and many-body
systems, where the simpler SRG-evolved wave functions are most
advantageous.

The practical pathway to few-body operators and applications to $A >2 $
is at present through the (Jacobi) harmonic-oscillator basis.
The successful SRG evolution of three-body forces (and higher in model
systems) in this basis is detailed in Refs.~\cite{Jurgenson:2008jp,Jurgenson:2009qs}.
The corresponding challenge is to evolve operators for $A=2$ and $A=3$
and then embed them (including induced contributions) in higher-$A$ spaces.
Procedures for carrying this out, including the need to boost some operators, 
will be
discussed in the sequel to this article \cite{Anderson:2010aa}. 
The calculation of transition matrix elements rather than ground-state
expectation values requires separate consideration.

Of particular interest will be the further study of SRG
operator factorization, which occurs when there is a scale separation
between the initial (unevolved) operator and the wave function momentum
scale, which is limited by $\lambda$.  
This factorization was shown to be a natural consequence of applying
the OPE to the unitary transformation. 
The extension to $A>2$ was previewed in an application of factorization for
momentum distributions of low-energy bound states, which provided an 
alternative interpretation to the commonly invoked role of short-range
correlations.
Work is in progress on the full realistic three-dimensional
calculations.

In closing, we reiterate that the favorable consequences of the SRG
is a specific realization of the more general observation that
the RG allows one
to focus on the most relevant degrees of freedom in a physical 
problem~\cite{Weinberg:1981qq}.
Thus, an evolution to low momentum for nuclear systems can be win-win
not only for the Hamiltonian and wave functions but for operators as well.
The SRG has some special advantages in practice because it uses operator 
flow equations
that can be applied in any convenient basis with 
a variety of options for tailoring the flow.
Furthermore, residual dependence on the flow parameter $s$ or $\lambda$
becomes a powerful tool for assessing approximations.
Different resolutions can lead to very different physical
interpretations and intuition; the 
RG has the great advantage of being able to connect the different
pictures.  

\begin{acknowledgments}
We
thank E. Jurgenson, A. Schwenk, and K. Wendt for useful comments and discussions.  We also thank E. Jurgenson for the use of his 1D few-body code.  This work was supported in part by the National Science Foundation under Grant Nos.~PHY--0653312 and PHY-0758125, and the UNEDF SciDAC Collaboration under DOE Grants DE-FC02-09ER41457 and DE-FC02-09ER41585. 

\end{acknowledgments}

%%%%%%%%%%%%%%%%%%%%%%%%% References %%%%%%%%%%%%%%%%%%%%%%%%%%%%%

\bibliographystyle{apsrev}
\bibliography{vlowk_refs}

\end{document}